\documentclass[11pt]{article}

\usepackage{graphicx}
\usepackage{amssymb}
\usepackage{amsmath}
\usepackage{times}
\usepackage{color}
\usepackage{cite}
\usepackage{authblk}

\usepackage[font=small,labelfont=bf]{caption}

\title{Binding and segregation of proteins in membrane adhesion: Theory, modelling, and simulations}
\author[1]{Thomas R.\ Weikl}
\author[2]{Jinglei Hu}
\author[1]{Batuhan Kav}
\author[3]{Bartosz R\'{o}\.{z}ycki}
\affil[1]{\footnotesize Max Planck Institute of Colloids and Interfaces, Theory and Bio-Systems, Potsdam, Germany}
\affil[2]{\footnotesize Kuang Yaming Honors School \& Institute for Brain Sciences, Nanjing University, Nanjing, China}
\affil[3]{\footnotesize Institute of Physics, Polish Academy of Sciences,  Warsaw, Poland}
\date{}                     
\date{}

\begin{document}

\maketitle

\tableofcontents

\section*{Abstract}
The adhesion of biomembranes is mediated by the binding of membrane-anchored receptor and ligand proteins. The proteins can only bind if the separation between apposing membranes is sufficiently close to the length of the protein complexes, which leads to an interplay between protein binding and membrane shape. In this article, we review current models of biomembrane adhesion and novel insights obtained from the models. Theory and simulations with elastic-membrane and coarse-grained molecular models of biomembrane adhesion indicate that the binding of proteins in membrane adhesion strongly depends on nanoscale shape fluctuations of the apposing membranes, which results in binding cooperativity. A length mismatch between protein complexes leads to repulsive interactions that are caused by membrane bending and act as a driving force for the length-based segregation of proteins during membrane adhesion.

\section*{Keywords}

cell adhesion; membrane bending energy;  membrane shape fluctuations; protein binding; binding cooperativity

\section{Introduction}

Cell adhesion and the adhesion of vesicles to the biomembranes of cells or organelles are central processes in immune responses, tissue formation, and cell signaling \cite{Alberts14}. These processes are mediated by a variety of receptor and ligand molecules that are anchored in the membranes. Biomembrane adhesion  is inherently multivalent -- the two apposing membranes contain numerous receptors and ligands and adhere {\em via} numerous receptor-ligand complexes, which ``clamp" the membranes together.  A striking feature are the largely different length scales from nanometers to micrometers involved in cell adhesion. Cells have typical diameters of tens of micrometers, but the separation of the apposing membranes in equilibrated cell adhesion zones is only about 15 to 40 nanometer \cite{Dustin00}. For typical concentrations of receptor-ligand complexes of the order of $100$ $\mu {\rm m}^{-2}$, the average separation between neighboring complexes is of the order of $\sqrt{1/100 \mu {\rm m}^{-2}}= 100$ nm. The membrane shape and shape fluctuations between neighboring receptor-ligand complexes are then dominated by the bending energy of the membranes with bending rigidity $\kappa$. On length scales of several hundred nanometers or microns, in contrast, the shape of cell membranes is dominated by the membrane tension $\sigma$ and by the membrane-connected actin cortex. The crossover from the dominance of the bending energy on small length scales to the dominance of membrane tension and actin cortex on large scales depends on the characteristic mesh size of the actin cortex of about 50 to 100 nm \cite{Morone06,Brown12,Etoc15} and on the characteristic length $\sqrt{\kappa/\sigma}$. For typical values of  the bending rigidity $\kappa$ of lipid membranes in the range from 10 to 40 $k_B T$ \cite{Nagle13,Dimova14} and typical tensions $\sigma$ of a few $\mu\text{N}/\text{m}$ \cite{Simson98,Popescu06,Betz09}, the characteristic length $\sqrt{\kappa/\sigma}$ adopts values between 100 and 400 nm.

Receptor and ligand molecules anchored in apposing membranes can only bind if the membrane separation at the site of these molecules is sufficiently close to the length of the receptor-ligand complex. Understanding biomembrane adhesion requires therefore an understanding of the interplay between (i) the specific binding of membrane-anchored receptor and ligand molecules and (ii) the membrane  shape as well as shape fluctuations between and around the receptor-ligand complexes. This review is focused on theory, modelling, and simulations of the interplay of protein binding and membrane shape in biomembrane adhesion. In chapter 2, we review various models of biomembrane adhesion that differ in their representation of the membranes and of the membrane-anchored receptors and ligands. In elastic-membrane models of biomembrane adhesion, the two adhering membranes are described as discretized elastic surfaces, and the receptors and ligands as molecules that diffuse along these surfaces.  In molecular models of biomembrane adhesion, the membrane lipids are captured as individual molecules,  either coarse-grained or atomistic, on the same level as the receptors and ligands.  Finally, multiscale modelling combines detailed molecular modelling of individual receptor-ligand complexes on short length scales with elastic-membrane modelling on large length scales. In chapter 3, we review insights on the binding equilibrium and binding constant of receptors and ligands in membrane adhesion zones obtained from elastic-membrane, coarse-grained molecular, and multiscale modelling.  A central result confirmed by recent experiments is the binding cooperativity of membrane-anchored receptors and ligands from membrane shape fluctuations on nanoscales \cite{Hu13,Krobath09,Steinkuhler19}.  In chapter 4, we consider the length-based segregation and domain formation of receptor-ligand complexes in membrane adhesion zones. Short receptor-ligand complexes tend to segregate from long receptor-ligand complexes or molecules because the membranes need to curve to compensate the length mismatch, which costs bending energy \cite{Weikl09,Weikl18}.  The article ends with a summary and outlook.

\begin{figure}[tp]
\begin{center}
\resizebox{\columnwidth}{!}{\includegraphics{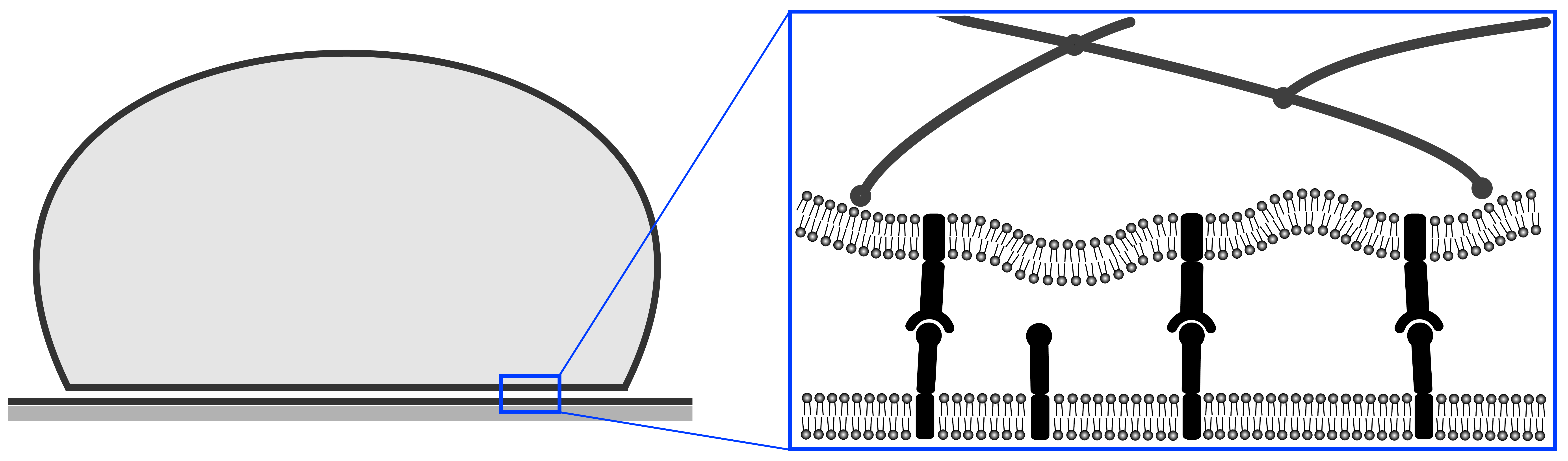}}
\end{center}
\caption{Cartoon of a cell that adheres to a substrate-supported membrane {\em via} receptor-ligand complexes.}
\label{figure_cartoon_adhesion}
\end{figure}
%

\section{Modelling biomembrane adhesion}

The adhesion of cells or vesicles leads to an adhesion zone in which receptors anchored in the cell or vesicle membrane interact with ligands anchored in the apposing membrane (see Fig.\ \ref{figure_cartoon_adhesion}). The challenge of modelling biomembrane adhesion is to model the interplay between the binding of these receptors and ligands and the membrane shape and shape fluctuations in the membrane adhesion zone. The theoretical models reviewed here differ in their spatial resolution and describe this interplay for smaller or larger segments of the adhesion zone and for mobile receptors that can diffuse within the membranes and are not strongly coupled to the cytoskeleton of an adhering cell. We first introduce the bending energy of adhering membranes, which is the basis of elastic-membrane and multiscale models of biomembrane adhesion.

\subsection{Bending energy of adhering membranes}
\label{subsection_bending_energy}

Biomembranes can be seen as thin elastic shells because their lateral extensions up to micrometers greatly exceed their thickness of a few nanometers.  In general, such thin shells can be deformed by shearing, stretching, or bending. Biomembranes are hardly stretchable and pose no shear resistance because of their fluidity. The dominant deformation of biomembranes therefore is bending. The bending energy depends on the local curvature of the membranes, which can be characterized by the two principal curvatures $c_1$ and $c_2$, or alternatively by the mean curvature $M = \frac{1}{2}(c_1 + c_2)$ and Gaussian curvature $K = c_1c_2$. The overall bending energy of a membrane is the integral \cite{Helfrich73}
\begin{equation}
E_\mathrm{be} = \int \left[2\kappa \Big(M - m\Big)^2 + \bar{\kappa} K\right] dA
\label{Ebe}
\end{equation}
over the membrane area $A$. The characteristic elastic parameters of the membrane are the bending rigidity $\kappa$ and the modulus of Gaussian curvature $\bar{\kappa}$. The spontaneous curvature $m$ reflects an intrinsic preference for bending, which can result from an asymmetry in either the composition of the two monolayers or in the concentration of solutes on both sides of the membranes \cite{Lipowsky13}. 

Our focus here is  on symmetric membranes with zero spontaneous curvature. For typical $\kappa$  values in the range from 10 to 40 $k_B T$ \cite{Nagle13,Dimova14} and typical  $\bar{\kappa}$ values between $-0.3 \kappa$ and $- \kappa$ of lipid membranes \cite{Hu12},  the bending energy (\ref{Ebe}) of symmetric membranes with  $m = 0$ is minimal in the planar state  \cite{Siegel04,Hu12}. Membrane shape changes and fluctuations relative to the planar minimum-energy state with zero bending energy can be described by the perpendicular deviation $h(x,y)$ out of a reference $x$-$y$-plane. For typically small angles between the membrane and the reference $x$-$y$-plane, the mean curvature $M$ can be approximated as $2 M \simeq   \Delta h(x,y) = \partial^2 h(x,y)/\partial x^2 + \partial^2 h(x,y)/\partial y^2$, and the bending energy is \cite{Safran94}
\begin{equation}
E_\mathrm{be} = \int {\textstyle\frac{1}{2}}\kappa \left(\Delta h\right)^2  \mathrm{d}x \,\mathrm{d}y
\label{Ebe-quasi-planar}
\end{equation}
because the integral over the Gaussian curvature $K$ in Eq.\ (\ref{Ebe}) is zero for all deviations and fluctuations that do not change the membrane topology of the planar state, according to the Gauss-Bonnet theorem. Contributions from boundary terms of the Gauss-Bonnet theorem vanish for appropriate choices of the boundary conditions, e.g. for periodic boundaries.  

The overall bending energy of two adhering membranes with negligible spontaneous curvature is the sum $E_\mathrm{be} = \int\left[ {\textstyle\frac{1}{2}}\kappa_1 \left(\Delta h_1\right)^2 + {\textstyle\frac{1}{2}}\kappa_2 \left(\Delta h_2\right)^2  \right] \mathrm{d}x \,\mathrm{d}y$ of the bending energies (\ref{Ebe-quasi-planar}) of the membranes. Here, $\kappa_1$ and $\kappa_2$ are the bending rigidities of the two membranes, and $h_1(x,y)$ and $h_2(x,y)$ are the deviation fields out of a reference $x$-$y$-plane. In analogy to the two-body problem,  a transformation of variables to the separation field $l = h_1-h_2$ and `center-of-mass' field $l_\text{cm}= \kappa_1 h_1 + \kappa_2 h_2$ allows to rewrite the overall bending energy as $E_\mathrm{be} = E_\text{ef}  + E_\text{cm}$ with the effective bending energy  \cite{Lipowsky88}
\begin{equation}
E_\mathrm{ef} = \int {\textstyle\frac{1}{2}}\kappa_\mathrm{ef}\left(\Delta l\right)^2 \mathrm{d}x \,\mathrm{d}y
\label{E-ef}
\end{equation}
for the separation field and the `center-of-mass'  energy $E_\text{cm} = \int {\textstyle\frac{1}{2}}(\kappa_1+\kappa_2)^{-1}$ $\left(\Delta l_\text{cm} \right)^2 \mathrm{d}x \,\mathrm{d}y$.  The effective binding rigidity in Eq.\ (\ref{E-ef}) is $\kappa_{\rm ef} = \kappa_1\kappa_2/(\kappa_1 + \kappa_2)$. Because the binding of receptors and ligands typically depends only on the local membrane separation $l$ \cite{Xu15}, the energy $E_\text{cm}$ is irrelevant for the specific binding of receptors and ligands.  If one of the membranes, e.g.~membrane 2, is a planar supported membrane, the effective bending rigidity $\kappa_{\rm ef}$ equals the rigidity $\kappa_1$ of the apposing membrane because the rigidity $\kappa_2$ of the supported membrane can be taken to be much larger than $\kappa_1$.

In simulations, elastic membranes are usually discretized. A discretization of the effective bending energy (\ref{E-ef}) of two adhering membranes can be achieved by discretizing the reference $x$-$y$-plane into a square lattice of lattice sites $i$ with lattice constant $a$ and local separations $l_i$. A discretized version of the effective bending energy (\ref{E-ef}) is \cite{Lipowsky96,Weikl06}
\begin{equation}
E_{\rm ef} = \frac{\kappa_\mathrm{ef}}{2 a^2} \sum_i \left( \Delta_{\rm d} l_i \right)^2
\label{E-ef-dis}
\end{equation}
with the discretized Laplacian $\Delta_{d}l_i =l_{i1}+l_{i2}+l_{i3}+l_{i4}-4 l_i$.  Here, $l_{i1}$ to $l_{i4}$ are the membrane separations at the four nearest-neighbor sites of site $i$ on the square lattice. The linear size $a$ of the membrane patches is typically chosen to be around 5 nm to capture the whole spectrum of bending deformations of the lipid membranes \cite{Goetz99}. 

\subsection{Elastic models of biomembrane adhesion}
\begin{figure}[tp]
\begin{center}
\resizebox{\columnwidth}{!}{\includegraphics{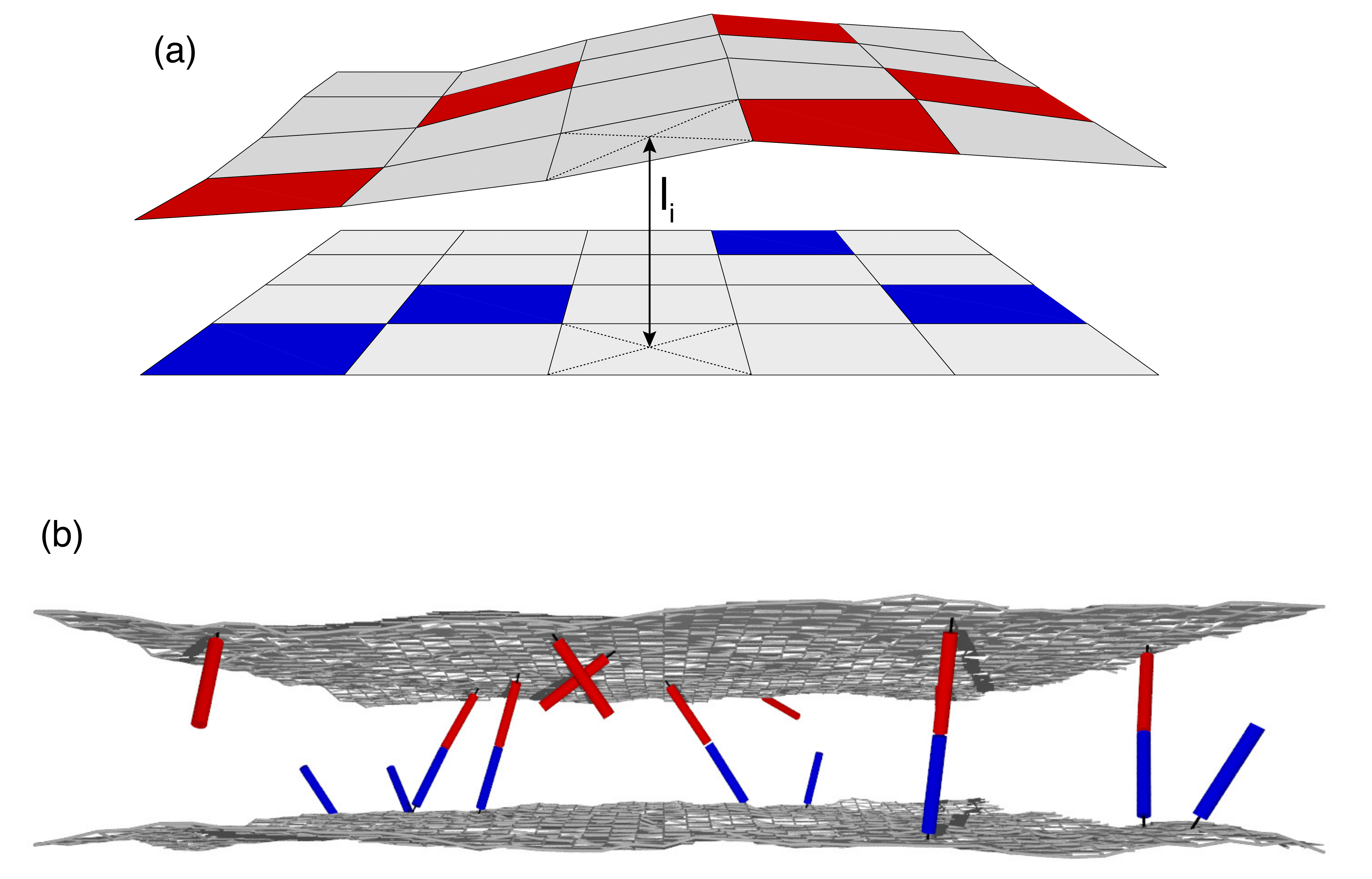}}
\end{center}
\caption{Elastic-membrane models of biomembrane adhesion. In both models, the membranes are described as discretized elastic surfaces (see section \ref{subsection_bending_energy}). In (a), the distribution of receptors and ligands in the quadratic patches of the discretized membranes is described by occupation numbers $n_i$  and $m_i$ that can adopt the values 0 and 1 \cite{Weikl01,Weikl06}. Here, $n_i=1$ indicates that a receptor is present in membrane patch $i$ of the upper membrane (red patches), and $m_i = 1$ indicates a ligand in patch $i$ of the lower membrane (blue patches). The values $n_i=0$ and $m_i = 0$ indicate patches without receptors and ligands (grey patches). Receptors and ligands at apposing membrane patches $i$ interact {\em via} a potential $V(l_i)$ that depends on the local separation $l_i$ of the patches (see Eq.\ (\ref{E_int})). The model can be generalized to membranes with different types of receptors and ligands (see Eq.\ (\ref{E_int_2})). In (b), the receptors and ligands are modelled as  anchored rods that diffuse continuously along the membranes and rotate around their anchoring points \cite{Xu15}. The binding and unbinding of the receptors and ligands is taken into account by  distance- and angle-dependent interactions of the binding sites, which are located at the tips of the receptors and ligands.
 }
\label{figure_elastic_membranes}
\end{figure}

The discretization of the reference plane in the effective bending energy (\ref{E-ef-dis}) implies a discretization of two adhering membranes into apposing pairs of nearly quadratic membrane patches. Discrete elastic models of biomembrane adhesion can be constructed by including receptor and ligand molecules in these membrane patches 
\cite{Lipowsky96,Weikl06,Weikl01,Weikl02a,Weikl04,Asfaw06,Tsourkas07,Reister08,Krobath09,Rozycki10,Krobath11,Bihr12,Li17,Knezevic18,Li19}. If the adhesion is mediated by a single type of receptor-ligand complexes, the distribution of receptors in one of the membranes can be described by the occupation numbers $n_i=1$ or $0$, which indicate whether a receptor is present or absent in patch $i$ of this membrane. In the same way, the distribution of ligands in the apposing membrane can be described by occupation numbers $m_i=1$ or $0$ (see Fig.\ \ref{figure_elastic_membranes}(a)). The specific interactions of the receptors and ligands can then be taken into account by the interaction energy \cite{Weikl01,Krobath09}
\begin{equation}
E_{\rm int}  = \sum_{i}  n_i m_i V(l_i)
\label{E_int}
\end{equation}
Here, $V(l_i)$ is the interaction potential of a receptor and a ligand that are present at the apposing membrane patches $i$, which implies $n_i=1$ and $m_i=1$. The interaction potential $V(l_i)$ depends on the local separation $l_i$ of the apposing patches. A simplified interaction potential is the square-well potential 
\begin{eqnarray}
V(l_i) &=& -U \mbox{~~for~~} l_o- l_{\rm we}/2 < l_i  < l_o + l_{\rm we}/2 \nonumber\\
  &=& 0 \mbox{~~otherwise} 
\label{square_well}
\end{eqnarray}
with the binding energy $U>0$, effective length $l_o$,  and binding range $l_{\rm we}$ of a receptor-ligand complex. For this interaction potential, a receptor and ligand are bound with energy $-U$ if the local separation $l_i$ of their apposing membrane patches is within the binding range $l_o \pm l_{\rm we}/2$.
  
The total energy of the model is the sum $E_{\rm tot} = E_{\rm be} +  E_{\rm int}$ of the effective bending energy (\ref{E-ef-dis}) of the membranes and the interaction energy (\ref{E_int}) of the receptors and ligands. The binding equilibrium of the biomembranes can be determined from the free energy  $F = -k_B T \ln Z $, where $Z$ is the partition function of the system, $k_B$ is Boltzmann's constant, and $T$ is the temperature. The partition function $Z$ is the sum over all possible membrane configurations, with each configuration weighted by the Boltzmann factor $\exp\left[-E_{\rm tot}  /k_B T \right]$. A membrane configuration is characterized by the local separations $\{l_i\}$ of the apposing membrane patches and by the distributions $\{m_i\}$  and $\{n_i\}$ of the receptors and ligands. In this model, the partial summation in the partition function $Z$ over all possible distributions $\{m_i\}$  and $\{n_i\}$  of receptors and ligands can be performed exactly, which leads to an effective adhesion potential. The effective adhesion potential $V_{\rm ef}(l_i)$ is a square-well potential  with the same binding range $l_{\rm we}$ as the receptor-ligand interaction (\ref{square_well}) and an effective potential depth $U_{\rm ef}$ that depends on the concentrations and binding energy $U$ of receptors and ligands \cite{Weikl01,Weikl06,Krobath09}: 
\begin{eqnarray}
V_{\rm ef}(l_i) &=& -U_{\rm ef} \mbox{~~for~~} l_o- l_{\rm we}/2 < l_i  < l_o + l_{\rm we}/2 \nonumber\\
  &=& 0 \mbox{~~otherwise} 
\label{Vef}
\end{eqnarray}
For typical concentrations of receptors and ligands up to hundred or several hundred molecules per square micrometer  in cell adhesion zones, the average distance between neighboring pairs of receptor and ligand molecules is much larger than the lattice spacing $a\simeq 5$ nm of the discretized membranes. For these small concentrations, the effective binding energy of the membranes is 
\begin{equation}
U_{\rm ef} \simeq k_B T \, K\,  [R] [L ]
\label{Uef}
\end{equation}
where $[R]$ is the area concentration of unbound receptors, $[L]$ is the area concentration of unbound ligands, and $K = a^2  e^{U/k_BT}$ is the binding constant for local separations inside the binding well \cite{Krobath09}. The binding equilibrium in the contact zone can therefore be determined from considering two membranes with the discrete elastic energy (\ref{E_int}) that interact {\em via} the effective adhesion potential (\ref{Vef}) with well depth $U_{\rm ef}$ and width $l_{\rm we}$. 

An important quantity is the fraction $P_b$ of membrane patches that are bound in the square well of the effective adhesion potential (\ref{Vef}). The bound membrane fraction $P_b$ can be determined in Monte Carlo simulations \cite{Weikl01,Weikl06} and continuously decays to zero when the effective binding energy $U_{\rm ef}$ approaches the critical binding energy $U_{\rm ef}^{c}$ for the unbinding of the membranes \cite{Lipowsky86}. The unbinding transition of the membranes results from an interplay of the effective adhesion (\ref{Vef}) and the steric repulsion of the membranes that is caused by thermally excited membrane shape fluctuations \cite{Helfrich78}. For effective binding energies smaller than $U_{\rm ef}^{c}$, the effective adhesion potential (\ref{Vef}) is not strong enough to hold the membranes together against their steric, fluctuation-induced repulsion.

Biomembrane adhesion {\em via} a single type of receptor-ligand bond occurs in experiments with a single type of ligand that binds to a single type of receptor in cells \cite{Dustin96,Zhu07,Tolentino08}, reconstituted vesicles \cite{Albersdoerfer97,Kloboucek99,Maier01,Smith06,Lorz07,Purrucker07,Smith08,Reister08,Fenz09,Monzel09,Streicher09,Smith10,Fenz17}, or cell-membrane-derived vesicles \cite{Steinkuhler19}. The ligands in these experiments are either anchored in substrate-supported membranes, or directly deposited on substrates. 
But cell adhesion is often mediated by several types of receptor-ligand complexes. The adhesion of T cells, natural killer (NK) cells, and B cells is mediated by receptor-ligand complexes of different lengths \cite{Davis04,Monks98,Grakoui99,Davis99,Batista01,Taylor17}. Cell adhesion {\em via} long and short receptor-ligand complexes has been investigated in experiments in which the cells adhere to supported membranes that contain two types of ligands \cite{Grakoui99,Mossman05,Milstein08,Huppa10,Axmann12,ODonoghue13}.  

Adhesion {\em via} two types of receptors and ligands can be modeled with occupation numbers $n_i$ and $m_i$ that adopt the values 0, 1, and 2, where $n_i=1$ and  $m_i=1$ indicate the presence of a receptor and ligand of type 1 at the apposing pair $i$ of membrane patches, and $n_i=2$ and $m_i=2$ indicate the presence of a receptor and ligand of type 2. The interaction energy
\begin{equation}
E_\text{int}= \sum_i \delta_{n_i,1}\delta_{m_i,1} V_1(l_i) + \delta_{n_i,2}\delta_{m_i,2} V_2(l_i)
\label{E_int_2}
\end{equation}
 of the membranes then includes two interaction potentials $V_1$ and $V_2$ for the two types of receptor-ligand complexes \cite{Weikl09,Asfaw06}. The Kronecker symbol $\delta_{j,k}$ in Eq.\ (\ref{E_int_2}) adopts the value $1$ for $j=k$ and the value $0$ for $j\neq k$. For square-well interaction potentials $V_1$ and $V_2$ with binding energies $U_1$ and $U_2$, effective lengths $l_1$ and $l_2$ of the complexes, and well widths  $l_\text{we}^{(1)}$ and  $l_\text{we}^{(2)}$, a summation over all possible distributions $\{m_i\}$  and $\{n_i\}$  of the receptors and ligands leads to an effective double-well adhesion potential if the length difference $|l_2 - l_1|$ of the receptor-ligand complexes is larger than the potential widths $l_\text{we}^{(1)}$ and  $l_\text{we}^{(2)}$. The two wells of this effective adhesion potential are located at $l_1$ and $l_2$, have the widths $l_\text{we}^{(1)}$ and  $l_\text{we}^{(2)}$, and the effective depths 
\begin{equation}
U_{1}^{\rm ef} \simeq k_B T \, K_1 \, [R_1] [L_1] 
\label{U1ef}
\end{equation}
and
\begin{equation}
U_{2}^{\rm ef} \simeq k_B T \, K_2 \, [R_2] [L_2] 
\label{U2ef}
\end{equation}
which depend on the concentrations $[R_1]$,  $[R_2]$, $[L_1]$, and $[L_2]$ of unbound receptors and ligands of type 1 and 2, and on the binding constants $K_1 = a^2 e^{U_1 / k_B T}$ and $K_2 = a^2 e^{U_2 / k_B T}$ of the receptors and ligands for local membrane separations inside their binding wells \cite{Weikl09}. 

In elastic-membrane models with the interaction energies  (\ref{E_int}) and (\ref{E_int_2}), the binding of receptors and ligands is described by interaction potentials of membrane patches that contain receptors and ligands. In a more recent elastic-membrane model, the receptors and ligands are modelled as anchored rods that diffuse continuously along the discretized membranes and rotate around their anchoring points (see Fig.\ \ref{figure_elastic_membranes}(b)) \cite{Xu15}. The total energy in this model is the sum $E_{\rm tot}=E_{\rm be}^{(1)} +E_{\rm be}^{(2)} +E_\text{int}+E_\text{anc}$ of the discretized bending energies (\ref{Ebe-quasi-planar}) of both membranes, the overall interaction energy $E_\text{int}$ of the rod-like receptors as ligands, and the overall anchoring energy $E_\text{anc}$. The overall anchoring energy is the sum of the anchoring potentials $V_{\rm anc}$ of the receptors and ligands. A simple anchoring potential is the harmonic potential  $V_\text{anc}=\frac {1}{2} k_a \theta_a^2$ with anchoring strength $k_a$ and anchoring angle $\theta_a$ of the receptors or ligands relative to the local membrane normal \cite{Xu15}. The overall interaction energy $E_\text{int}$ in this model is the sum over the distance- and angle-dependent interaction potentials of the binding sites located at the tips of the receptors and ligands. 

In both types of elastic-membrane models illustrated in Fig.\ \ref{figure_elastic_membranes}, the receptors and ligands are modelled as individual molecules that can bind and unbind. In other elastic-membrane models, receptors and ligands have been described {\em via} concentration fields and not as individual molecules \cite{Komura00,Bruinsma00,Qi01,Chen03,Raychaudhuri03,Coombs04,Shenoy05,Wu06,Zhang08a,Atilgan09}, or receptor-ligand bonds have been treated as constraints on the local membrane separation \cite{Zuckerman95,Krobath07,Speck10,Weil10,Dharan15}.

\subsection{Molecular models of biomembrane adhesion}
\label{subsection_molecular_models}

\begin{figure}[tp]
\begin{center}
\resizebox{\columnwidth}{!}{\includegraphics{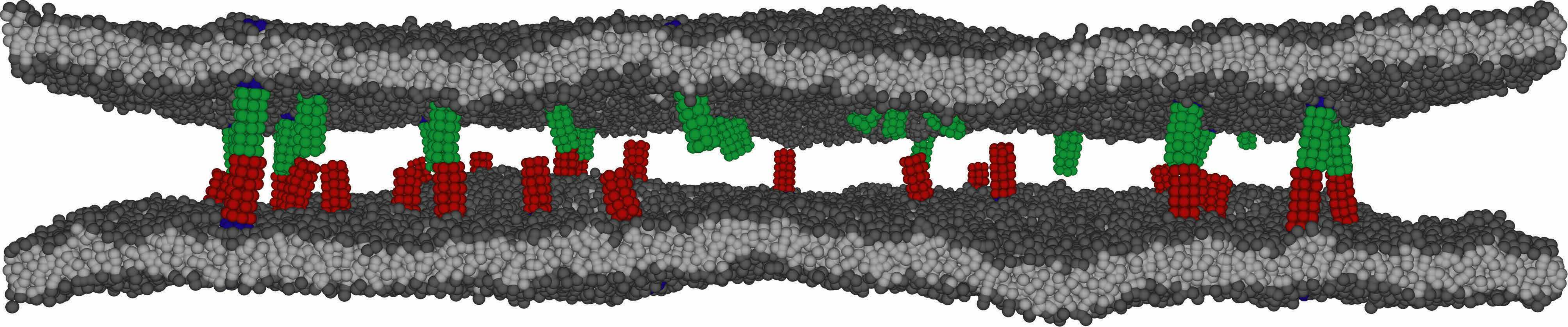}}
\end{center}
\caption{
Snapshot from a molecular dynamics simulation of a coarse-grained molecular model of biomembrane adhesion with transmembrane receptors and ligands \cite{Hu13,Hu15}. Each of the membranes contains 22136 lipids, 25 receptors or ligands, and has an area of $120\times 120$ $\text{nm}^2$. The lipid molecules consist of three hydrophilic head beads (dark gray) and two hydrophobic chains with four beads each (light grey) \cite{Goetz98,Shillcock02,Grafmuller07,Grafmuller09}. The transmembrane receptors and ligands consist of 84 beads (12 layers of 7 beads) arranged in a cylindrical shape and have hydrophobic anchors that are embedded in the lipid bilayer. The interaction domain of the receptors (green) and ligands (red) consists of six layers of hydrophilic beads, with an interaction bead or `binding site' located in the center of the top layer of seven beads. The transmembrane anchors of the receptors or ligands mimic the transmembrane segments of membrane proteins and are composed of four  layers of hydrophobic lipid-chain-like beads (not visible) in between two  layers of lipid-head-like beads (blue). For clarity, the water beads of the model are not displayed in the snapshot. The boundaries of the simulation box (not shown) are periodic. 
}
\label{figure_coarse}
\end{figure}

In molecular models of biomembrane adhesion, both the membrane lipids and the anchored receptors and ligands are modelled as individual molecules. Coarse-grained modelling and simulations have been widely used to investigate the self-assembly \cite{Goetz98,Shelley01,Marrink04,Shih06}, fusion \cite{Marrink03,Shillcock05,Grafmuller07,Grafmuller09,Smirnova10,Risselada11}, and lipid domains \cite{Illya06,Risselada08,Meyer10,Apajalahti10,Bennett13} of membranes as well as the diffusion \cite{Gambin06,Guigas06}, aggregation \cite{Reynwar07}, and curvature generation \cite{Arkhipov08,Simunovic13,Braun14,Dasgupta18} of membrane inclusions and membrane proteins. In the coarse-grained molecular model of biomembrane adhesion \cite{Hu13,Hu15} shown in Fig.\ \ref{figure_coarse}, the lipid molecules consist of three hydrophilic head beads (dark gray) and two hydrophobic chains with four beads each (light grey) \cite{Goetz98,Shillcock02,Grafmuller07,Grafmuller09}. The receptors and ligands are composed of an interaction domain that protrudes out of the membrane (green, red), and a membrane anchor (not visible). The interaction domain consists of hydrophilic beads arranged in a cylindrical shape, with an `interaction bead' as binding site at the center of the tip. This interaction domain is either rigidly connected to a cylindrical transmembrane anchor that contains hydrophobic beads, or is flexibly connected to a lipid molecule \cite{Hu15}. The specific binding of the receptors and ligands is modeled {\em via} a distance- and angle-dependent binding potential between two interaction beads at the tip of the molecules.  The binding potential has no barrier to ensure an efficient sampling of binding and unbinding events of receptors and ligands in our simulations. The average binding times of receptors and ligands in this potential are of the order of 10 $\mu s$. Total simulation times of tens of milliseconds then lead to typically thousands of binding and unbinding events, which allows to determine the binding equilibrium and kinetics with high precision \cite{Hu13,Hu15}. The length and time scales of this coarse-grained model have been estimated by comparing to experimental data for the bilayer thickness and lipid diffusion constant of fluid membranes.

\begin{figure}[tp]
\begin{center}
\resizebox{\columnwidth}{!}{\includegraphics{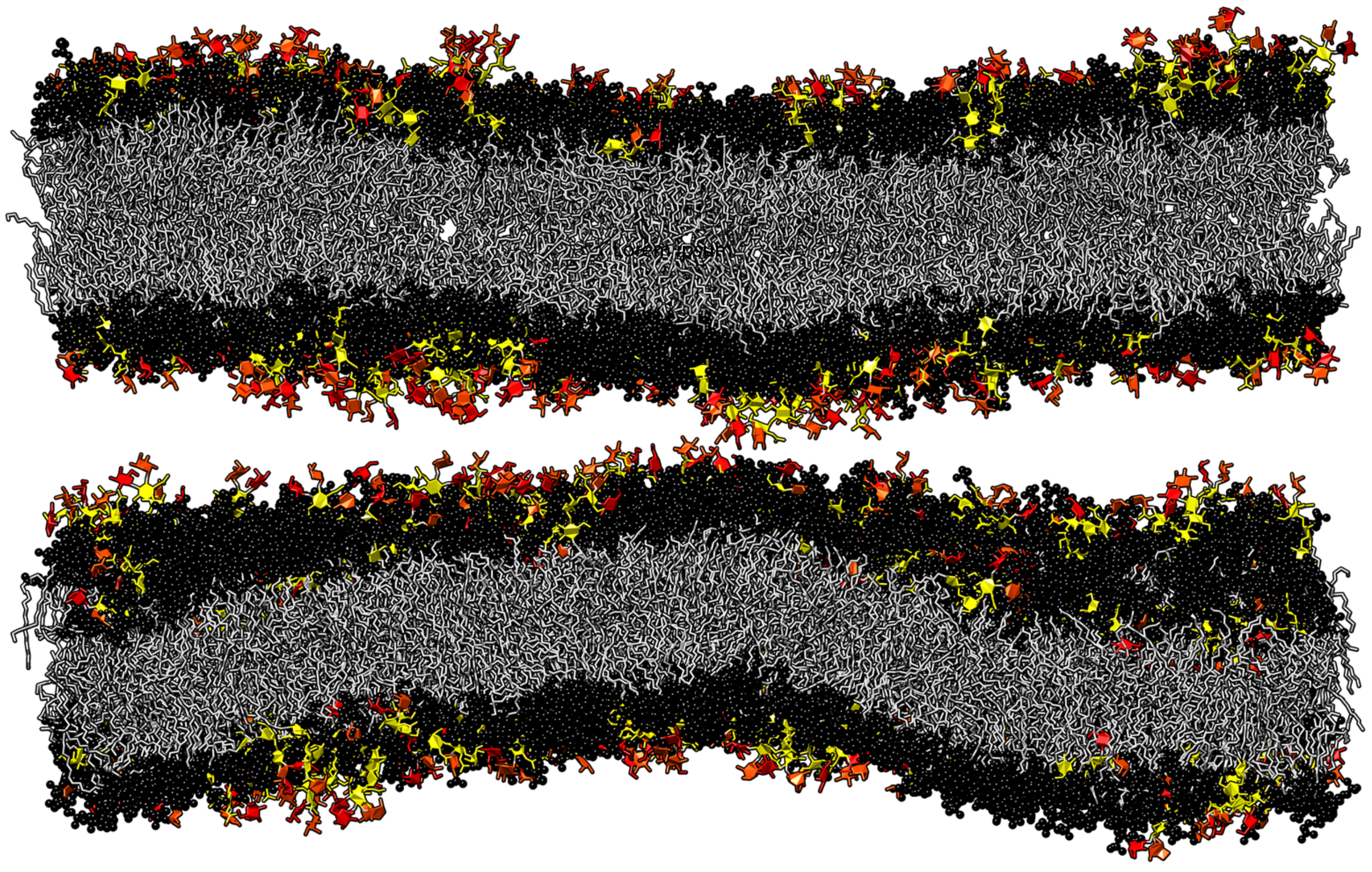}}
\end{center}
\caption{Snapshot from an atomistic simulation of two apposing membranes that interact {\em via} lipid-anchored Lewis-X (LeX) saccharides \cite{Kav19}. The fucose and galactose rings at the tip of LeX are represented in red and orange. The remaining three sugar rings that are connected to the lipid tails are given in yellow. Lipid heads and tails are shown in black and gray, respectively. 
}
\label{figure_atomistic}
\end{figure}

Atomistic simulation trajectories of protein binding and biomembrane adhesion are currently limited to typical timescales of microseconds \cite{Plattner17,Paul17,Kav19}, which is orders of magnitude smaller than the binding times of receptor and ligand proteins in membrane adhesion \cite{Huppa10,Huang10,Robert11,Axmann12,ODonoghue13}.  The binding equilibrium of proteins in membrane adhesion systems is therefore beyond the scope of current atomistic simulations. But in some systems, membrane adhesion can also be mediated by glycolipids \cite{Schneck11} with binding times of nanoseconds in the atomistic simulations of Fig.\ \ref{figure_atomistic} \cite{Kav19}. Simulation times of one microsecond are therefore sufficient to reach binding equilibrium. In the atomistic simulations of Fig.\ \ref{figure_atomistic}, the two membranes are composed of 1620 lipids  and 180 glycolipids each and have an area of about $25 \times 25$ nm$^2$. The glycolipids contain five sugar rings that are connected to lipid tails \cite{Schneck11}. 

Besides reaching the binding equilibrium of receptors and ligands, a challenge for coarse-grained and atomistic simulations of biomembrane adhesion is to equilibrate the thermal shape fluctuations of the membranes. The equilibration time depends on the lateral correlation length  $\xi_\parallel$ of the membranes \cite{Lipowsky89,Seifert94}, which in turn depends on the concentration $[{\rm RL}]$ of the receptor-ligand bonds \cite{Xu15}, and on the lateral size of the membrane in simulations of small membrane systems with few bonds \cite{Hu13}. In the membrane systems of Figs.\ \ref{figure_coarse} and \ref{figure_atomistic}, the correlation length is smaller than the lateral extension of the membranes and, thus, dominated by the concentration $[{\rm RL}]$ of bonds, which constrain the membrane shape fluctuations. Membrane shape fluctuations lead to variations in the local separation of the membranes, which can be quantified by the relative roughness $\xi_\perp$.  The relative roughness $\xi_\perp$ is the standard deviation of the variations in the local separation. For fluid membranes, the thermal roughness  is proportional to the correlation length $\xi_\parallel$ \cite{Lipowsky95}, which in turn is proportional to the average distance $1/\sqrt{[{\rm RL}]}$ between neighboring bonds for sufficiently large membrane adhesion systems. Theory and simulations indicate the approximate scaling relation \cite{Xu15}
\begin{equation}
\xi_\perp \simeq 0.2 \sqrt{(k_B T / \kappa_{\rm ef})} \Big /
\sqrt{[{\rm RL}]}
\label{relative_roughness_scaling}
\end{equation}
 with the effective bending rigidity $\kappa_{\rm ef} = \kappa_1\kappa_2/(\kappa_1 + \kappa_2)$ introduced in Eq.\ (\ref{E-ef}). The scaling relation (\ref{relative_roughness_scaling}) holds for roughnesses smaller than the length of the receptor-ligand bonds. For such roughnesses, the fluctuation-induced, steric repulsion of the membranes is negligible. For a typical concentration $[{\rm RL}]\simeq 100/\mu{\rm m}^{2}$ of receptor-ligand bonds in cell adhesion and for typical values of the bending rigidities $\kappa_1$ and $\kappa_2$ of lipid membranes between 10 $k_B T$ and 40 $k_B T$  \cite{Nagle13,Dimova14}, the relative membrane roughness $\xi_\perp$ attains values between 3 nm and 6 nm according to Eq.\ (\ref{relative_roughness_scaling}). These calculated roughness values are smaller than the lengths of receptor-ligand bonds in cell adhesion, which are typically between 15 and 40 nanometers \cite{Dustin00}.

\subsection{Multiscale modelling}
\label{subsection_multiscale_modelling}

\begin{figure}[htp]
\begin{center}
\resizebox{0.9\columnwidth}{!}{\includegraphics{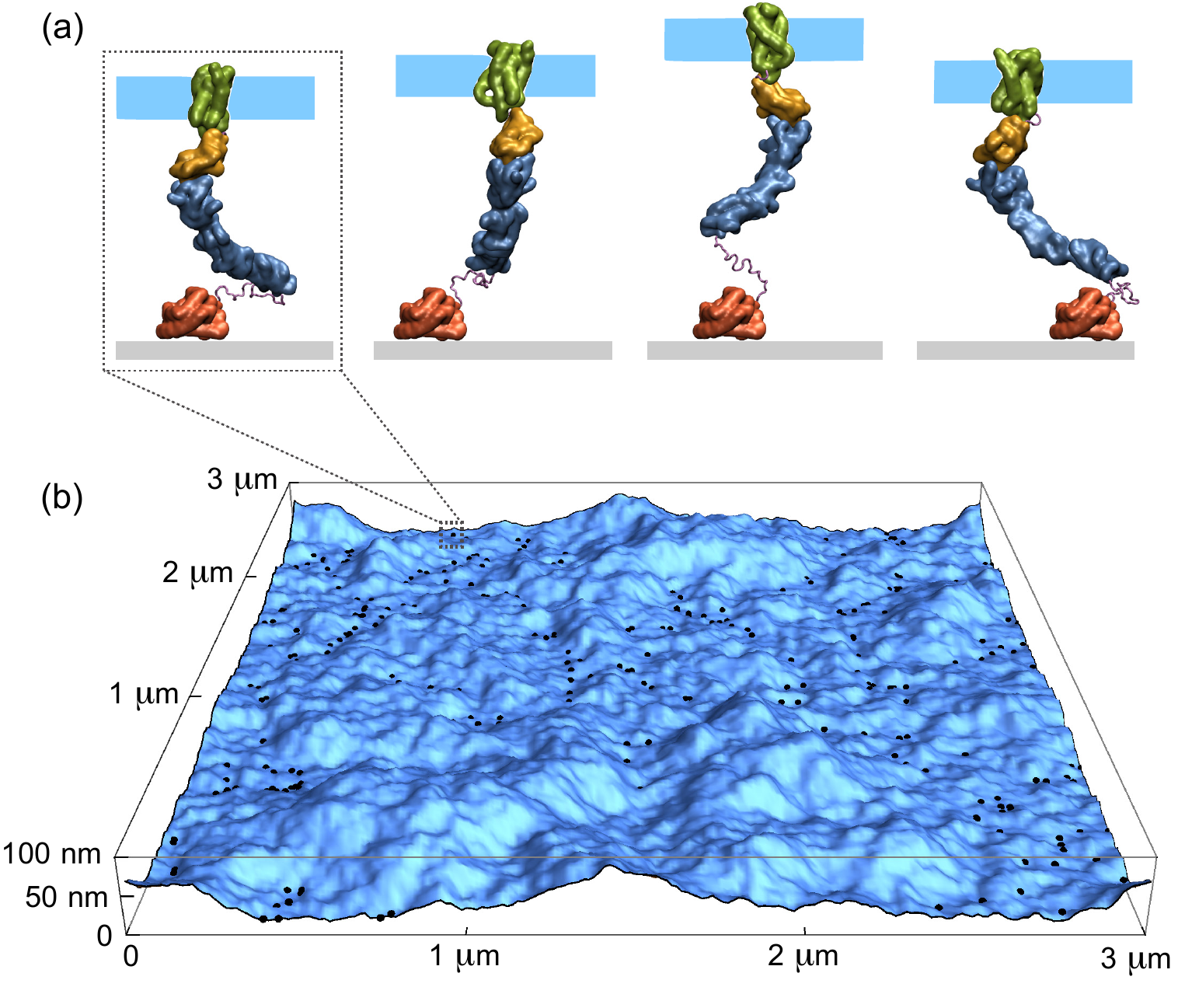}}
\end{center}
\caption{Multiscale modelling of membrane adhesion mediated by CD47-SIRP$\alpha$ complexes \cite{Steinkuhler19}: (a) Snapshots from coarse-grained molecular simulations of a CD47-SIRP$\alpha$ complex. The separation between the membrane patch (light blue) and the substrate (grey) varies in the simulations mainly due to conformational changes of the unstructured linker that covalently connects SIRP$\alpha$ (blue) to substrate-bound GST proteins (red). The transmembrane and interaction domain of CD47 are shown in green and yellow. (b) Snapshot from a simulation of an adhering membrane segment with area $3 \times 3$ $\mu\rm{m}^2$ for the concentration $[\rm{RL}] = 30$ $\mu\text{m}^{-2}$ of CD47-SIRP$\alpha$  complexes (black dots) at the parameter value $\Delta l = 4$ nm of the repulsive membrane-substrate interactions. The complexes are modelled as elastic springs with a spring constant determined from the molecular simulations. The repulsive interactions between the protein layer on the substrate and the membrane are taken into account by allowing only local separations $l > l_0 – \Delta l$ between the membrane patches and the substrate, where $l_0$ is the preferred separation of the CD47-SIRP$\alpha$ complexes for binding. }
\label{figure_multiscale_model}
\end{figure}

The detailed modelling of protein binding in membrane adhesion is complicated by the relatively long binding times of proteins from  milliseconds to seconds \cite{Huppa10,Huang10,Robert11,Axmann12,ODonoghue13} and by the relatively large lateral membrane sizes of tens or hundreds of nanometers that are required to capture the full spectrum of membrane shape fluctuations up to the correlation length of the adhering membranes. A promising approach is multiscale modelling, i.e.\  a combination of molecular modelling of proteins on short length scales and elastic modelling of membranes on longer length scales. Fig.\ \ref{figure_multiscale_model} illustrates a multiscale-modelling approach for the adhesion of vesicle membranes with the marker-of-self protein CD47 to SIRP$\alpha$ proteins immobilized on a surface. This approach combines (i) coarse-grained molecular modelling and simulations of a single CD47-SIRP$\alpha$ complex (see Fig.\ \ref{figure_multiscale_model}(a)) to determine the effective spring constant of the complex, and (ii) simulations of large elastic membrane segments adhering {\em via} CD47-SIRP$\alpha$ complexes, modelled as elastic springs (see Fig.\ \ref{figure_multiscale_model}(b)). 

The molecular simulations of the CD47-SIRP$\alpha$ complex are based on the Kim-Hummer model,  in which large multi-domain proteins are divided into rigid domains and unstructured linkers that connect these domains \cite{Kim08}. CD47 is divided into a rigid trans-membrane domain (green) and a rigid binding domain (yellow), which are connected by a rather short linker segment and an additional harmonic bond that mimics a disulfide bridge. The rigid SIRP$\alpha$ interaction domain (blue) is connected to a surface-bound GST domain by a relatively long linker. In the molecular simulations, the CD47 transmembrane domain can rotate in a plane parallel to the substrate surfaces but keeps its orientation relative to the substrate in order to mimic the membrane embedding of this domain. The separation between the CD47 transmembrane domain and the substrate varies in the simulations and depends on the molecular architecture and flexibility of the CD47-SIRP$\alpha$ complex. The flexibility of the complex is dominated by the rather long unstructured linker segment that connects SIRP$\alpha$ and GST, which leads to the standard deviation $\sigma \simeq 1.2$ nm in the separation of the CD47 transmembrane domain and the substrate. This standard deviation of the membrane-substrate separation corresponds to an effective spring constant $k_S=k_BT/\sigma^2$ of the complex. The mean value $l_0 \simeq 17.2$ nm of the separation is the preferred membrane-substrate separation of the complex.

The elastic-membrane simulations of Fig.\ \ref{figure_multiscale_model}(b) are based on the discretized bending energy described in section \ref{subsection_bending_energy}. The large adhering membrane segments in these simulations are discretized into quadratic patches with linear size  $a\simeq 15$ nm. This larger patch size allows to equilibrate the largescale membrane shape fluctuations, which dominate the membrane roughness,  for the rather small complex concentrations of $[\rm{RL}] = 30$ $\mu\text{m}^{-2}$ of CD47-SIRP$\alpha$. Membrane patches that contain CD47-SIRP$\alpha$ complexes (black dots) are bound to the substrate {\em via} a harmonic potential with the preferred separation $l_0$ and spring constant $k_S$ of the complexes obtained from the molecular simulations.  Repulsive interactions between the protein layer on the substrate and the membrane are taken into account by allowing only local separations $l > l_0 – \Delta l$ between the membrane patches and the substrate, with $\Delta l > 0$. The CD47-SIRP$\alpha$ complexes in the simulations are mobile and diffuse along the membrane by hopping from patch to patch. In the experiments, the mobility of complexes results from unbinding and rebinding of the relatively few CD47 proteins to the many SIRP$\alpha$ proteins on the substrate \cite{Steinkuhler19}.

\section{Binding of membrane-anchored receptor and ligand proteins}

\subsection{General theory for the binding constant of proteins in membrane adhesion}
\label{section_K2D}

Adhesion processes of cells depend sensitively on the binding affinity of the membrane-anchored receptor and ligand molecules that mediate adhesion. Binding affinities are typically quantified by binding equilibrium constants. For soluble receptor and ligand molecules that are free to diffuse in three dimensions, the binding constant $K_{\rm 3D} = [{\rm RL}]_{\rm 3D} /[{\rm R}]_{\rm 3D} [{\rm L}]_{\rm 3D} $ is fully determined by the binding free energy of the receptor-ligand complex RL. Here,  $[{\rm RL}]_{\rm 3D}$, $[{\rm R}]_{\rm 3D}$, and $[{\rm L}]_{\rm 3D}$ are the volume concentrations of the complex RL and of the unbound receptors R and ligands L. A corresponding quantity for membrane-anchored receptors and ligands is $K_{\rm 2D} = [{\rm RL}]_{\rm 2D} /[{\rm R}]_{\rm 2D} [{\rm L}]_{\rm 2D}$ where $[{\rm RL}]_{\rm 2D}$, $[{\rm R}]_{\rm 2D}$, and $[{\rm L}]_{\rm 2D}$ are the area concentrations of the complex RL and of the unbound receptors R and ligands L  \cite{Dustin01,Orsello01} (see Fig.\ 6). In contrast to $K_{\rm 3D}$, the quantity $K_{\rm 2D}$ is not fully determined by the receptors and ligands, but depends also on properties of the membranes. For example, $K_{\rm 2D}$ is zero if the membrane separation is significantly larger than the length of the receptor-ligand complex, because RL complexes cannot form at such large membrane separations. 

\begin{figure}[tp]
\begin{center}
\resizebox{\columnwidth}{!}{\includegraphics{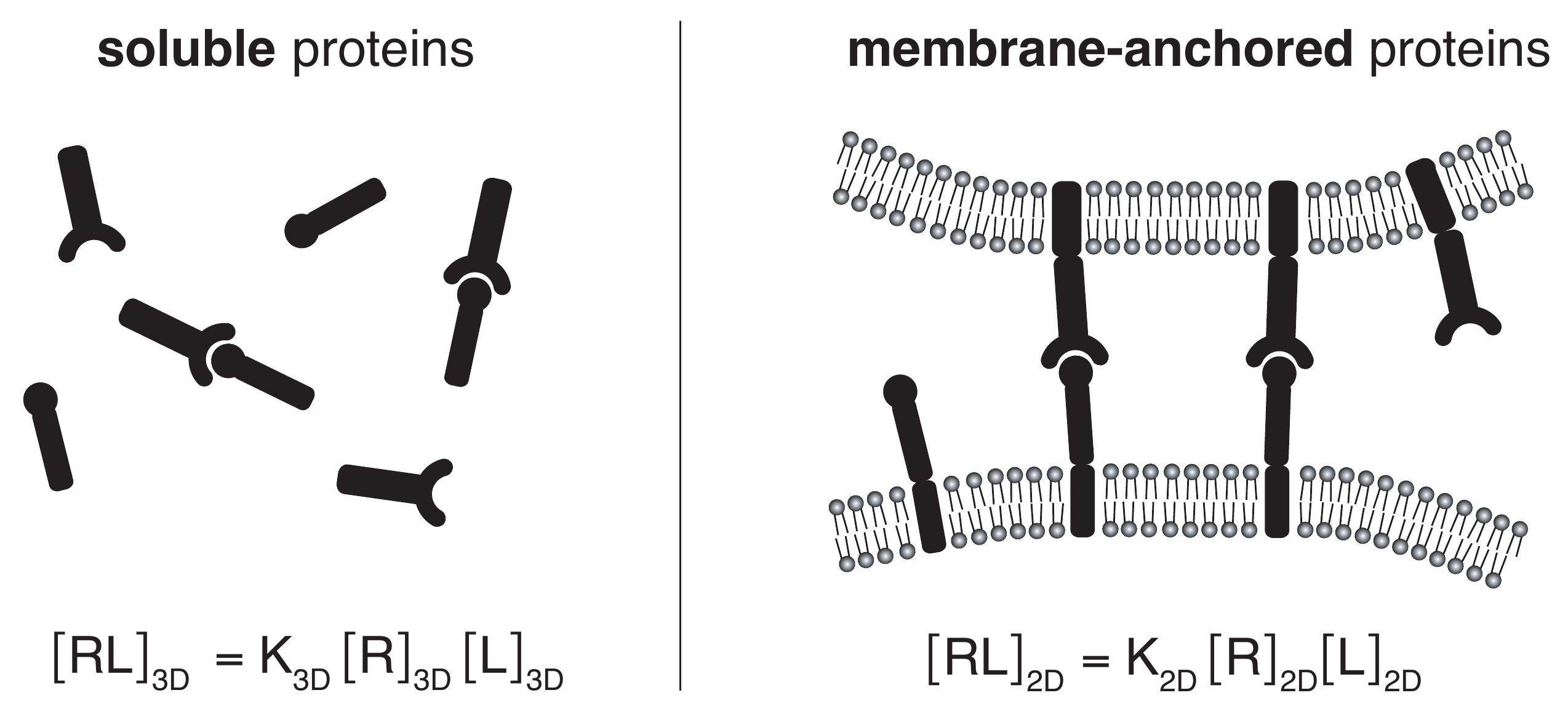}}
\end{center}
\caption{The binding constant $K_{\rm 3D}$ of soluble receptor and ligand molecules is determined by the binding free energy of the molecules. The corresponding quantity $K_{\rm 2D}$ of membrane-anchored receptor and ligand molecules, in contrast, does not only depend on properties of the molecules, but depends also on the membrane separation because the receptors and ligands can only bind if this separation is sufficiently close to the length of the receptor-ligand complexes. 
}
\label{figure_binding_constants}
\end{figure}

A membrane-anchored receptor can only bind to an apposing membrane-an- chored ligand if the local membrane separation $l$ at the site of the receptor and ligand is within an appropriate range. This local separation $l$ of the membranes varies -- along the membranes, and in time -- because of thermally excited membrane shape fluctuations. Measurements of $K_{\rm 2D}$ imply an averaging in space and time over membrane adhesion regions and measurement durations. Theory and simulations with the elastic-membrane model of Fig.\ \ref{figure_elastic_membranes}(b) and the coarse-grained molecular model of Fig.\ \ref{figure_coarse} indicate that this averaging  can be expressed as \cite{Xu15,Hu15}
\begin{equation}
K_{\rm 2D} = \int k_{\rm 2D}(l) P(l) {\rm d}l 
\label{K2Dav}
\end{equation}
where $k_{\rm 2D}(l)$ is the binding equilibrium constant of the receptors and ligands as a function of the local membrane separation $l$, and $P(l)$ is the distribution of local membrane separations that reflects the spatial and temporal variations of $l$. The binding constant function $k_{\rm 2D}(l)$ is determined by molecular properties of the membrane-anchored receptors and ligands and is, thus, the proper equivalent of the binding konstant $K_{\rm 3D}$ of soluble receptors and ligands. However, the set of molecular features that affect $k_{\rm 2D}(l)$ is rather large and includes the lengths, flexibility, and membrane anchoring of the receptors and ligands, besides the interactions at the binding site \cite{Xu15,Hu15}. A softer membrane anchoring, for example, leads to smaller values of  $k_{\rm 2D}(l)$  because softly anchored receptors and ligands have more rotational entropy to loose during binding, compared to more rigidly anchored receptors and ligands. The anchoring elasticity depends on whether the receptors and ligands are anchored {\em via} lipids or trans-membrane protein domains, and on the length and flexibility of unstructured linker segments that connect the extracellular domains of the receptors and ligands to these membrane anchors. If the adhesion is mediated by a single type of receptor-ligand complex with a length that is larger than the relative roughness $\xi_\perp$ of the fluctuating membranes from thermally excited shape fluctuations, the distribution $P(l)$ of the local separation can be approximated by the Gaussian function \cite{Hu15,Xu15}
\begin{equation}
P(l) \simeq \exp\left[-(l-\bar l)^2/2\xi_\perp^2\right]/(\sqrt{2\pi} \xi_\perp)
\label{Pl}
\end{equation}
The mean of the distribution $P(l)$ is the average separation $\bar{l}$ of the membranes, and the standard deviation of this distribution is the relative roughness $\xi_\perp$.

\begin{figure}[htp]
\begin{center}
\resizebox{\columnwidth}{!}{\includegraphics{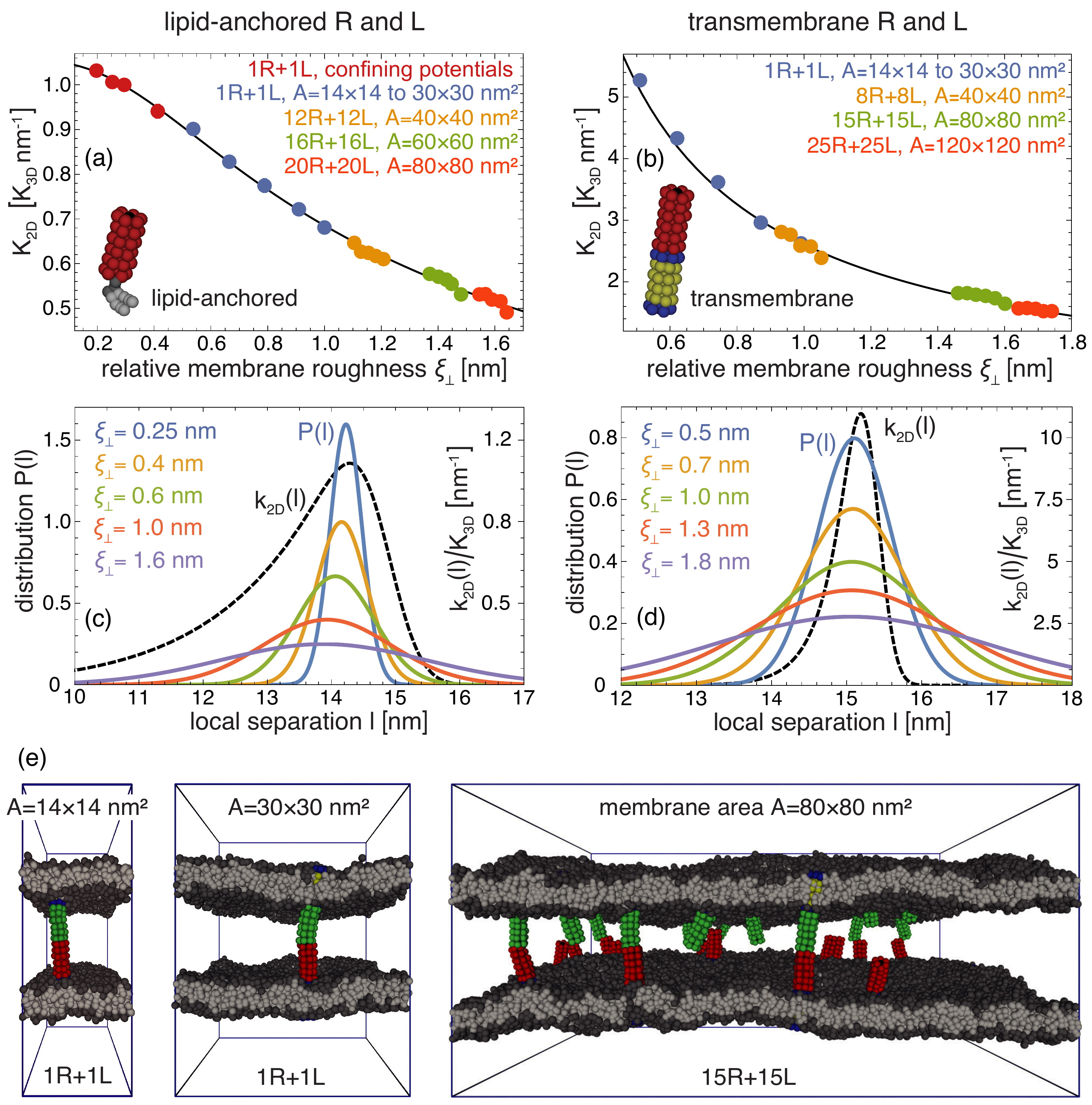}}
\end{center}
\caption{ (a) and (b) $K_{\rm 2D}$ for lipid-anchored and transmembrane receptors and ligands versus 
relative membrane roughness $\xi_\perp$ at the `optimal' average separation for binding \cite{Hu15}. The MD data points result from a variety of membrane systems with transmembrane and lipid-anchored receptors and ligands of a coarse-grained molecular model of biomembrane adhesion.  In these systems, the area $A$ of the two apposing membranes ranges from $14 \times 14$ nm$^2$ to $120 \times 120$ nm$^2$, and the number of receptors R and ligands L varies between 1 and 25 (see figure legends). The units of $K_{\rm 2D}$ are $K_{\rm 3D}/$nm where  
$K_{\rm 3D} \simeq 157$ nm$^3$ is the binding constant of soluble variants of the receptors and ligands without membrane anchors. The full lines in (a) and (b) represents fits based on Eqs.\ (\ref{K2Dav}) and (\ref{Pl}) with functions $k_{\rm 2D}(l)$ from modelling the receptor-ligand bonds as harmonic springs that can tilt \cite{Xu15,Hu15}. These functions $k_{\rm 2D}(l)$ are shown in (c) and (d), together with the distributions $P(l)$ of local separations for various relative membrane roughnesses $\xi_\perp$. (e) Snapshots from MD simulations of three different membrane systems with transmembrane receptors and ligands.
}
\label{figure_DPD_data}
\end{figure}

Fig.\ \ref{figure_DPD_data} illustrates the modelling of simulation results for $K_{\rm 2D}$  based on Eq.\ (\ref{K2Dav}). The data points in Fig.\ \ref{figure_DPD_data}(a) and (b) result from MD simulations with the coarse-grained molecular model of Fig.\ \ref{figure_coarse}. In these simulations, $K_{\rm 2D}$ has been measured for a variety of membrane systems that differ in membrane area, in the number of receptors and ligands, or in the membrane potential \cite{Hu15}, at `optimal' average membrane separations $\bar l$ at which $K_{\rm 2D}$ is maximal. These optimal average separations correspond to average separations in equilibrated adhesion zones of cell, because maxima in $K_{\rm 2D}$ correspond to minima in the adhesion free energy \cite{Hu13,Xu15}. In the MD simulations, the average separation is determined by the number of water beads between the membranes and is, thus, constant. The relative membrane roughness $\xi_\perp$ depends on the area $L_x \times L_y$ of the membranes because the periodic boundaries of the simulation box suppress membrane shape fluctuations with wavelength larger than $L_x/2 \pi$. In membrane systems with several anchored receptors and ligands, the roughness is affected by the number of receptor-ligand bonds because the bonds constrain the membrane shape fluctuations. For the small numbers of receptors and ligands in the MD simulations, the binding constants can be determined from the times spent in bound and unbound states \cite{Hu13,Hu15}.  

With increasing size of the membrane systems, the relative roughness $\xi_\perp$ strongly increases, while $K_{\rm 2D}$  strongly decreases. The decrease of $K_{\rm 2D}$ with increasing roughness $\xi_\perp$ can be understood from Eq.\ (\ref{K2Dav}). An increase in $\xi_\perp$ implies a broadening of $P(l)$, which leads to smaller values of the integral $\int k_{\rm 2D}(l) P(l) {\rm d}l$ at the optimal average separation of binding. In Fig.\ \ref{figure_DPD_data}(c) and (d), the distributions  $P(l)$ at the optimal average separation are illustrated for several values of the membrane roughness $\xi_\perp$. The function $k_{\rm 2D}(l)$ for the lipid-anchored receptors and ligands in Fig.\ \ref{figure_DPD_data}(c) is significantly broader than $k_{\rm 2D}(l)$ for the  transmembrane receptors and ligands in Fig.\ \ref{figure_DPD_data}(d). The functions  $k_{\rm 2D}(l)$ are asymmetric because the receptor-ligand complexes can tilt at local separations $l$ smaller than the `preferred' separation $l_0$ at which the functions are maximal, but need to stretch at local separations larger than $l_0$.

Because  $K_{\rm 2D}$ is difficult to measure in adhesion experiments, an important question is how $K_{\rm 2D}$ is related to the binding equilibrium constant $K_{\rm 3D}$ of soluble variants of the receptors and ligands that lack the membrane anchors \cite{Dustin01,Orsello01,Krobath09,Wu11,Leckband12,Zarnitsyna12,Hu13,Wu13,Xu15,Hu15}. The binding constant $K_{\rm 3D}$ can be quantified with standard experimental methods \cite{Schuck97,Rich00,McDonnell01}. In Fig.\ \ref{figure_DPD_data}, $K_{\rm 2D}$ is given in units of $K_{\rm 3D}/$nm where $K_{\rm 3D}\simeq 157$ nm$^3$ is determined by the binding potential of the coarse-grained model \cite{Hu13,Hu15}. For the rod-like receptors and ligands of this coarse-grained molecular model and of  the elastic-membrane model of Fig.\ \ref{figure_elastic_membranes}(b), the ratio $k_{\rm 2D}(l)/K_{\rm 3D}$ can be calculated from the loss of rotational and translational entropy  during binding \cite{Xu15,Hu15}:
\begin{equation}
k_{\rm 2D}(l)/K_{\rm 3D} \simeq  \sqrt{8\pi}  \frac{A_b}{V_b} \frac{\Omega_{\rm RL}(l)}{\Omega_{\rm R}\Omega_{\rm L}}
\label{k2DoverK3D}
\end{equation}
Here,  $\Omega_{\rm R}$, $\Omega_{\rm L}$, and $\Omega_{\rm RL}(l)$ are the rotational phase space volumes of the unbound receptors R, unbound ligands L, and bound receptor-ligand complex RL relative to the membranes, and $A_b$ and $V_b$ are the translational phase space area and translational phase space volume of the bound ligand relative to the receptor in 2D and 3D. The rotational phase space volume $\Omega_{\rm RL}(l)$ depends on the local separation $l$ of the membranes and can be calculated by modelling the membrane-anchored complex as a harmonic spring that can tilt \cite{Xu15,Hu15}. The Eqs.\ (\ref{K2Dav}), (\ref{Pl}), and (\ref{k2DoverK3D}) provide a general relation between $K_{\rm 2D}$  and $K_{\rm 3D}$. The ratio $K_{\rm 2D}/K_{\rm 3D}$ has units of an inverse length, but depends on several length scales: the average separation $\bar{l}$ and relative roughness $\xi_\perp$ of the membranes, the width of the function $k_{\rm 2D}(l)$, and the ratio $V_b/A_b$ in Eq.\ (\ref{k2DoverK3D}). The ratio  $V_b/A_b$ is a characteristic length for the binding interface of the receptor-ligand complex and can be estimated as the standard deviation of the binding-site distance in the direction of the complex  \cite{Xu15}.

\subsection{Cooperative protein binding from membrane shape fluctuations on nanoscales}
\label{section_cooperative_binding}

\begin{figure}[htp]
\begin{center}
\resizebox{\columnwidth}{!}{\includegraphics{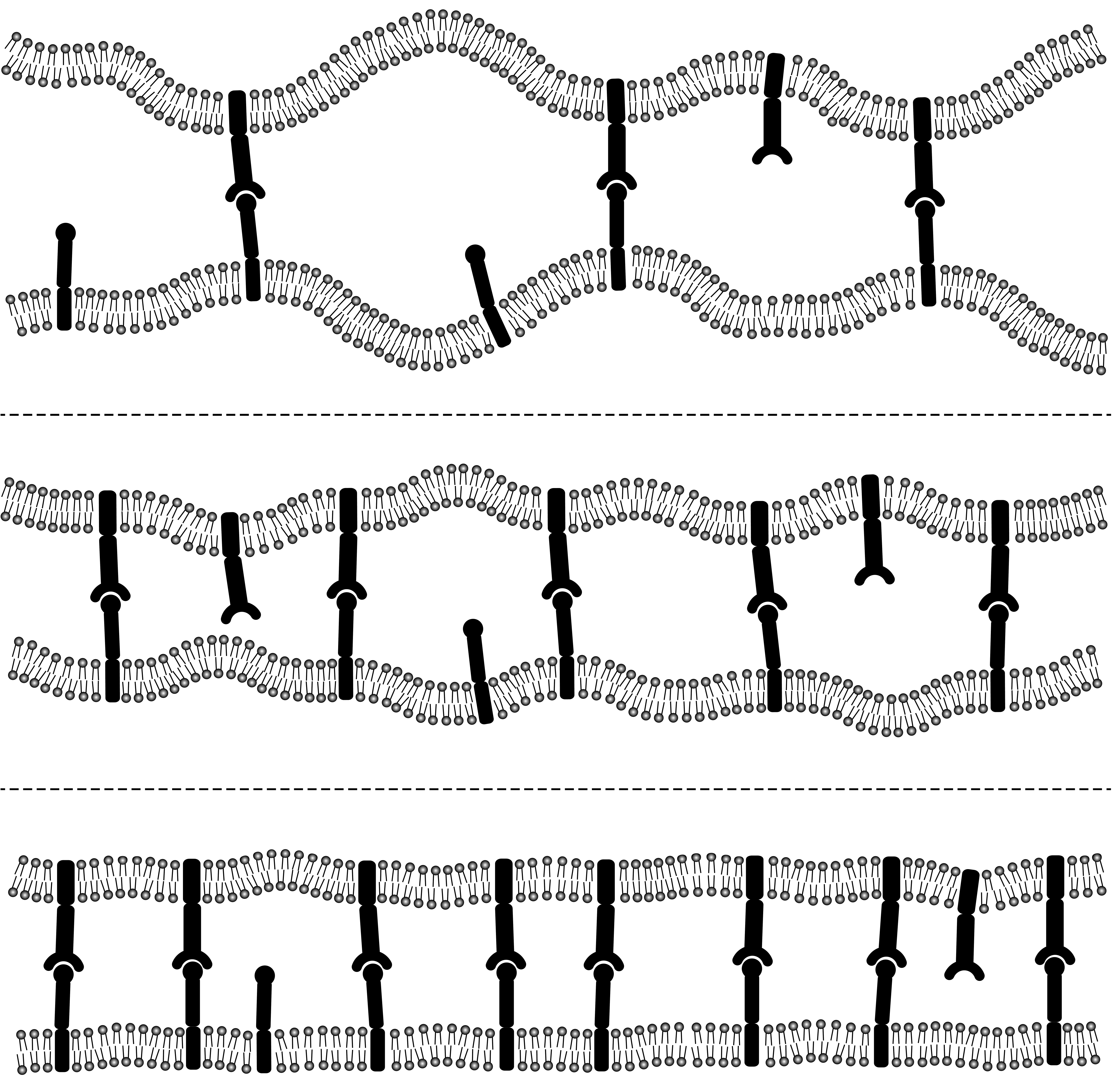}}
\end{center}
\caption{The relative roughness of adhering membranes decreases with increasing concentration of receptor–ligand bonds  because the bonds constrain membrane shape fluctuations (see also Eq.\ref{relative_roughness_scaling}). The binding constant $K_{\rm 2D}$ of the receptors and ligands in turn increases with decreasing roughness (see Eq.\ (\ref{binding_cooperativity})), which leads to cooperative binding.
}
\label{figure_cooperativity}
\end{figure}

\begin{figure}[tp]
\begin{center}
\resizebox{\columnwidth}{!}{\includegraphics{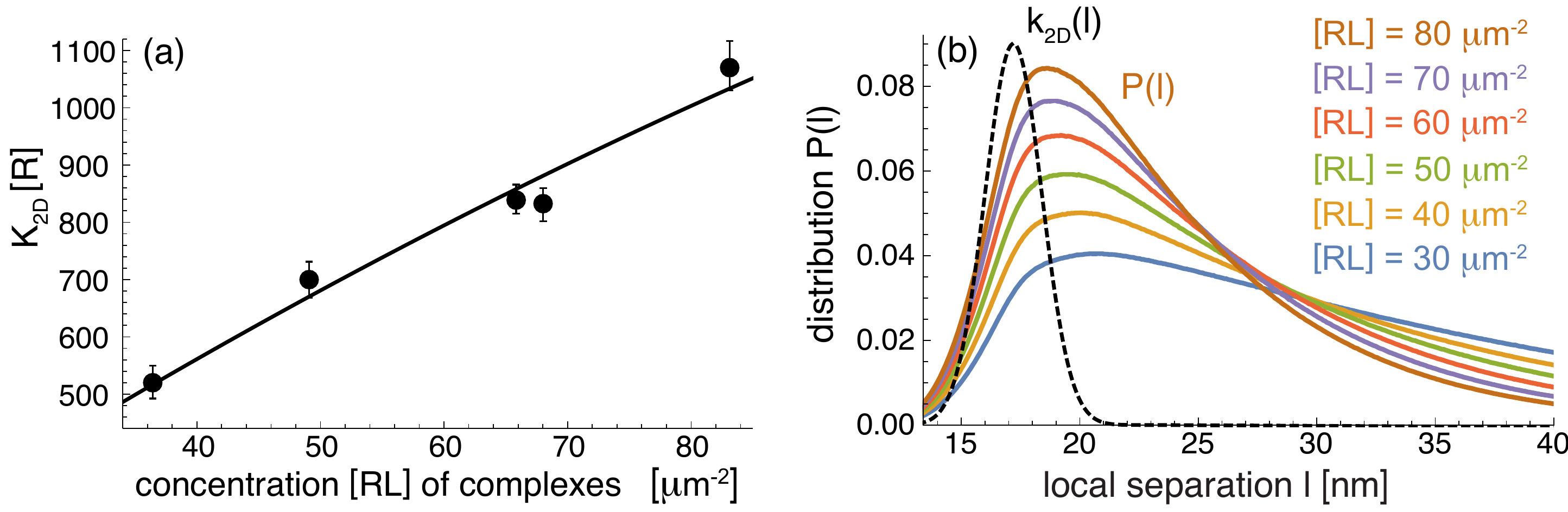}}
\end{center}
\caption{(a) $K_{\rm 2D}$ of CD47-SIRP$\alpha$ complexes as a function of complex concentration $[{\rm RL}]$ from experiments (data points) and multiscale modelling (line) based on Eq.\ (\ref{K2Dav}). (b) Shape of the function $k_{\rm 2D}(l)$ determined from coarse-grained molecular simulations of a single CD47-SIRP$\alpha$ complex (see Fig.\ \ref{figure_multiscale_model}(a)), and distributions $P(l)$ of local membrane separations from elastic-membrane simulations with different complex concentrations $[{\rm RL}]$ (see Fig.\ \ref{figure_multiscale_model}(b)). In the elastic-membrane simulations, the complexes are modelled as harmonic springs with a spring constant determined from the coarse-grained molecular simulations. Repulsive interactions between the protein layer on the substrate and the membrane in these simulations are taken into account by allowing only local separations $l > l_0 – \Delta l$ where $l_0\simeq 17.2$ nm is the preferred separation of the complexes for binding. The parameter value for the repulsive membrane-substrate interactions here is $\Delta l = 4$ nm. The maximum value  $K_{\rm max} =(378 \pm 8)/[{\rm R}]$ of the function $k_{\rm 2D}(l)$ is fitted to the data in (a), where $[{\rm R}]$ is the concentration of unbound SIRP$\alpha$ on the substrate. 
}
\label{figure_CD47_data}
\end{figure}

As discussed in the previous section, two important length scales in the general theory for $K_{\rm 2D}$ are the relative membrane roughness $\xi_\perp$ and the average membrane separation $\bar{l}$, i.e.\ the standard deviation and mean of the distribution $P(l)$ of local membrane separations. A third important length scale is the width of the function $k_{\rm 2D}(l)$ in Eq.\ (\ref{K2Dav}), which reflects the membrane confinement exerted by a single receptor-ligand complex.  Coarse-grained molecular modelling indicates that the standard deviation of the function $k_{\rm 2D}(l)$ is about 1 nm or less, depending on the anchoring of the receptors and ligands \cite{Steinkuhler19,Hu15} (see Figs.\ \ref{figure_DPD_data} and \ref{figure_CD47_data}). The relative membrane roughness $\xi_\perp$ depends on the concentration $[{\rm RL}]$ of receptor-ligand complexes, which `clamp' the membranes together and thus suppress membrane shape fluctuations, and on the bending rigidities  $\kappa_1$ and $\kappa_2$ of the adhering membranes. The scaling relation (\ref{relative_roughness_scaling}) leads to $\xi_\perp$ values between 3 nm and 6 nm for a typical concentration $[{\rm RL}]\simeq 100/\mu{\rm m}^{2}$ of receptor-ligand bonds in cell adhesion and for typical values of the bending rigidities $\kappa_1$ and $\kappa_2$ of lipid membranes  between 10 $k_B T$ and 40 $k_B T$ \cite{Nagle13,Dimova14}. For such relative membrane roughnesses, the distribution $P(l)$ of local membrane separations is significantly broader than the function $k_{\rm 2D}(l)$. Eq.\ (\ref{K2Dav}) with the Gaussian distribution $P(l)$ of Eq.\ (\ref{Pl}) can then be approximated as \cite{Xu15}
\begin{equation}
K_{\rm 2D} \simeq P(l_o) \int k_{\rm 2D}(l)  {\rm d}l \sim 1/\xi_\perp
\label{K2Dapprox} 
\end{equation}
for average separations $\bar l$ close to the preferred separation $l_0$ of the bonds. Eq.\ (\ref{K2Dapprox}) implies an inverse proportionality between $K_{\rm 2D}$ and $\xi_\perp$ at the optimal average separation $\bar l$ for binding, which has been first observed in coarse-grained MD simulations with transmembrane receptors and ligands \cite{Hu13}. For these transmembrane receptors and ligands, the  standard deviation of the function $k_{\rm 2D}(l)$ in Fig.\ \ref{figure_DPD_data}(d) is about 0.37 nm and, thus, smaller than all roughness values of Fig.\ \ref{figure_DPD_data}(b), and the black line in Fig.\ \ref{figure_DPD_data}(b) is $K_{\rm 2D}/K_{\rm 3D} \simeq 2.6 / \xi_\perp$ \cite{Weikl16}.  For the lipid-anchored receptors and ligands of Fig.\ \ref{figure_DPD_data}, the  standard deviation of the function $k_{\rm 2D}(l)$ in Fig.\ \ref{figure_DPD_data}(c)  is about 1.4 nm, and the inverse proportionality of Eq.\ (\ref{K2Dapprox}) holds for roughness values larger than about 2 nm. Together with the scaling relation (\ref{relative_roughness_scaling}) between the relative roughness $\xi_\perp$ and the concentration $[\rm RL]$ of receptor-ligand bonds, the inverse proportionality of $K_{\rm 2D}$ and $\xi_\perp$ leads to the law of mass action \cite{Hu13,Krobath09}:
\begin{equation}
 [{\rm RL}] =K_{\rm 2D} [{\rm R}][{\rm L}] \sim  [{\rm R}]^2[{\rm L}]^2
 \label{binding_cooperativity}
\end{equation}
The quadratic dependence of the bond concentration $[{\rm RL}]$ on the concentrations $[{\rm R}]$ and $[{\rm L}]$ of unbound receptors and ligands in Eq.\ (\ref{binding_cooperativity}) indicates cooperative binding. The binding cooperativity results from a smoothening of the membranes with increasing bond concentration $[{\rm RL}]$, which facilitates the binding of additional receptors and ligands (see Fig.\ \ref{figure_cooperativity}). 

The cooperative binding of receptors and ligands in membrane adhesion has been recently confirmed in experiments in which CD47 proteins in giant vesicles generated from cell membranes bind to SIRP$\alpha$ proteins immobilized on a planar substrate \cite{Steinkuhler19}. The experimental data points in Fig.\ \ref{figure_CD47_data}(a) show that $K_{\rm 2D}$ increases with increasing concentration $[{\rm RL}]$ of receptor-ligand bonds.  The multiscale modelling illustrated in Fig.\ \ref{figure_multiscale_model} confirms that this experimentally observed increase in $K_{\rm 2D}$ results from a smoothening of the membranes with increasing  $[{\rm RL}]$. In the multiscale modelling approach, the shape of the function $k_{\rm 2D}(l)$ is determined from coarse-grained simulations of a single CD47-SIRP$\alpha$ complex (see Fig.\ \ref{figure_multiscale_model}(a)). The distribution of membrane-substrate separations at the site of complex obtained from these simulations is equivalent to the shape of $k_{\rm 2D}(l)$, because of $k_{\rm 2D}(l) \sim \exp\left(-\Delta G_{\rm 2D}(l)/k_B T\right)$ where $\Delta G_{\rm 2D}(l)$ is the binding free energy of the complex \cite{Xu15}, and because the Boltzmann factor $\exp\left(-\Delta G_{\rm 2D}(l)/k_B T\right)$ determines the distributions of local separations at the site of the complex.  The coarse-grained molecular simulations indicate that the variations of the local separation at the site of the complex are dominated by the rather long unstructured linker segment that connects SIRP$\alpha$ to the substrate-bound GST domain in Fig.\ \ref{figure_multiscale_model}(a). The resulting distribution of local separations is approximately Gaussian, with a standard deviation $\sigma \simeq 1.2$ nm that corresponds to the width of $k_{\rm 2D}(l)$. In the elastic-membrane simulations of Fig.\ \ref{figure_multiscale_model}(b),  bound CD47-SIRP$\alpha$ complexes are therefore modelled as harmonic springs with effective spring constant $k_S=k_BT/\sigma^2$. The distributions $P(l)$ for the large adhering membrane segments of these simulations are shown in Fig.\ \ref{figure_CD47_data}(b) for various concentrations $[{\rm RL}]$ of CD47-SIRP$\alpha$ complexes. The distributions $P(l)$ are asymmetric because the width of these distributions, the relative membrane roughness  $\xi_\perp$, is larger than the separation $\Delta l = 4$ nm between the membrane and the protein layer in the simulations.  Steric interactions between the protein layer on the substrate and the membrane are taken into account by allowing only local separations $l > l_0 – \Delta l$ where $l_0$ is the preferred membrane-substrate separation of CD47-SIRP$\alpha$ complex. The steric interactions lead to a fluctuation-induced repulsion of the membrane from the protein layer on the substrate. Because of this fluctuation-induced repulsion, the average membrane-substrate separation $\bar{l}$, i.e.\ the mean of $P(l)$, is larger than the preferred membrane-substrate separation of CD47-SIRP$\alpha$ complex, which is $l_0\simeq 17.2$ nm according to the coarse-grained molecular simulations of Fig.\ \ref{figure_multiscale_model}(a). With increasing concentration $[{\rm RL}]$, the distributions $P(l)$ of Fig.\ \ref{figure_CD47_data}(b) become narrower and shift towards $l_0$, which both leads to larger values of $K_{\rm 2D}$ according to Eq.\ (\ref{K2Dav}) due to a larger overlap of $P(l)$ with $k_{\rm 2D}(l)$. The line in Fig.\ \ref{figure_CD47_data}(a) results from the distributions in (b) with the fit parameter $K_{\rm max} =(378 \pm 8)/[{\rm R}]$ for the maximal value of  $k_{\rm 2D}(l)$, where $[{\rm R}]$ is the concentration of unbound SIRP$\alpha$ on the substrate. Similar lines are obtained for values of the modelling parameter $\Delta l$ between 1 nm and 5 nm  \cite{Steinkuhler19}.

\section{Adhesion-induced domain formation of membrane-anchored proteins}

\subsection{Length-based segregation of receptor-ligand complexes}
\label{section_length_based}

The adhesion of T cells and other immune cells is mediated by receptor-ligand complexes of different lengths, which tend to segregate into domains of long and short complexes \cite{Davis04,Monks98,Grakoui99,Davis99,Batista01,Taylor17}. Long and short receptor-ligand complexes in membrane adhesion zones repel each other because the membranes need to curve to compensate the length mismatch (see Fig.\ 10). The strength of this curvature-mediated repulsion and segregation depends both on the length difference and the concentrations of the receptor-ligand complexes. 

Calculations and simulations with the elastic-membrane model of Fig.\ \ref{figure_elastic_membranes} with the interaction energy (\ref{E_int_2}) for short complexes ${\rm R}_1{\rm L}_1$ of length $l_1$ and long complexes  ${\rm R}_2{\rm L}_2$ of length $l_2$  indicate that the curvature-mediated segregation of the complexes is stable for bond concentrations 
\begin{equation}
[{\rm R}_1{\rm L}_1] = [{\rm R}_2{\rm L}_2] > \frac{c\, k_B T}{\kappa_\text{ef}(l_2 -l_1)^2}
\label{segregation}
\end{equation}
with the numerical prefactor $c = 0.65\pm 0.15$ \cite{Weikl18,Asfaw06,Weikl09}. Here, $\kappa_\text{ef} = \kappa_1\kappa_2/(\kappa_1+\kappa_2)$ is the effective bending rigidity of the two membranes with rigidities $\kappa_1$ and $\kappa_2$ (see section \ref{subsection_bending_energy}). In equilibrium, the concentration $[{\rm R}_1{\rm L}_1]$ of the short complexes in domain 1 is equal to the concentration $[{\rm R}_2{\rm L}_2]$ of the long complexes in domain 2. For complex concentrations  $[{\rm R}_1{\rm L}_1]$ and $[{\rm R}_2{\rm L}_2]$ smaller than the  critical concentration $c \,k_B T/\kappa_\text{ef}(l_2 -l_1)^2$ of Eq.~(\ref{segregation}), the domains are unstable because the entropy of mixing of the bonds then dominates over the curvature-mediated repulsion. 

\begin{figure}[tp]
\begin{center}
\resizebox{\columnwidth}{!}{\includegraphics{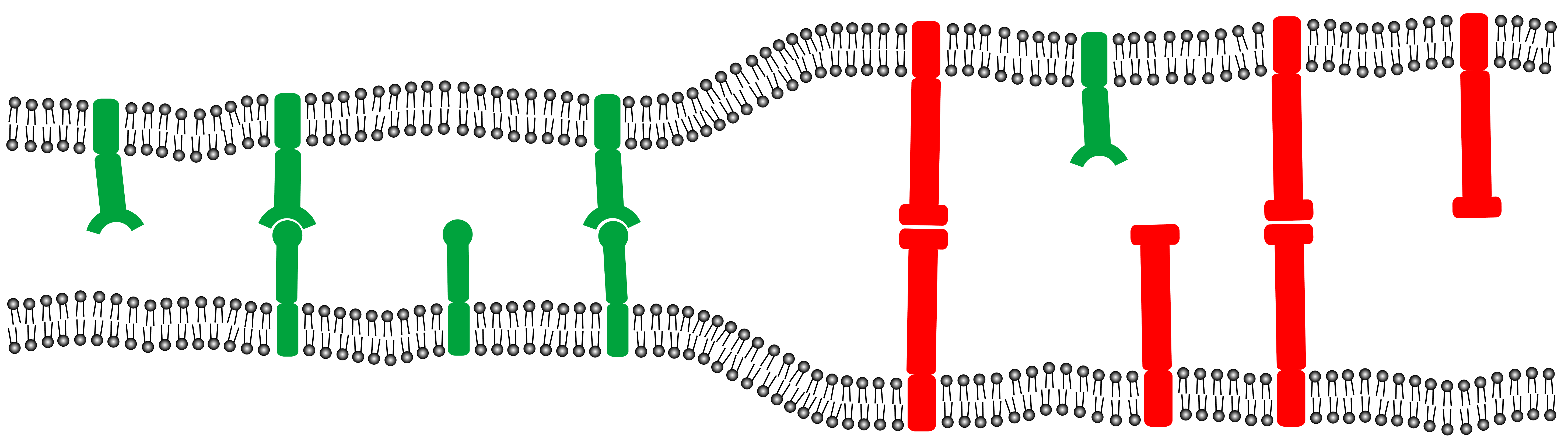}}
\end{center}
\caption{Long and short receptor-ligand complexes repel each other because the membranes need to curve to compensate the length mismatch. This curvature-mediated repulsion leads to domains of long and short receptor-ligand complexes for sufficiently large complex concentrations.}
\label{figure_segregation}
\end{figure}

The derivation of Eq.\ (\ref{segregation}) is based on the effective double-well potential of the elastic-membrane model with well depths $U_{1}^{\rm ef}$ and  $U_{1}^{\rm ef}$ that depend on the concentrations of the receptors and ligands and on their binding constants $K_1$ and $K_2$ for membrane separations within the potential wells (see Eq.\ (\ref{U1ef}) and (\ref{U2ef})). Elastic membranes interacting {\em via} this double-membrane potential exhibit stable domains for $U_\text{ef}^{(1)} l_\text{we}^{(1)} =U_\text{ef}^{(2)} l_\text{we}^{(2)} > c_a (k_BT)^2/\kappa (l_2 -l_1)$ with $c_a = 0.225\pm 0.020$  where $l_\text{we}^{(1)}$ and  $l_\text{we}^{(2)}$ are the widths the two wells \cite{Asfaw06}.  For typical relative membrane roughnesses in the domains larger than the well widths $l_\text{we}^{(1)}$ and  $l_\text{we}^{(2)}$, the bond concentrations $[{\rm R}_1{\rm L}_1] = c_b (\kappa/k_BT) \left(l_\text{we}^{(1)} K_1   [{\rm R}_1][{\rm L}_1]\right)^2$ and $[{\rm R}_2{\rm L}_2] = c_b (\kappa/k_BT) \left(l_\text{we}^{(2)} K_2   [{\rm R}_2][{\rm L}_2]\right)^2$  with $c_b = (13\pm 1)$ are proportional to the square of concentrations of the unbound receptors and ligands \cite{Krobath09,Weikl09} (see also Eq.~(\ref{binding_cooperativity}). Together, these equations lead to Eq.\ (\ref{segregation}) \cite{Weikl18}.

In T cell adhesion zones, nanoclusters and domains of short TCR-MHCp complexes  and long integrin complexes form at sufficiently large concentrations of MHCp \cite{Grakoui99,Campi05,Yokosuka05,Mossman05,Choudhuri10}. For the preferred membrane separations $l_1 \simeq 15$ nm and $l_2\simeq 40$ nm of the TCR-MHCp and integrin complexes \cite{Dustin00} and for typical membrane bending rigidities between 10 and 40 $k_B T$, the critical concentrations of Eq.\ (\ref{segregation}) for curvature-mediated segregation into domains vary between 25 and 200 complexes per square micron. These bond concentrations are smaller than the experimentally measured concentrations of TCR and integrin complexes \cite{Grakoui99}, which indicates that the curvature-mediated repulsion of these complexes is sufficiently strong to drive the domain formation. 

\begin{figure}[t]
\begin{center}
\resizebox{\columnwidth}{!}{\includegraphics{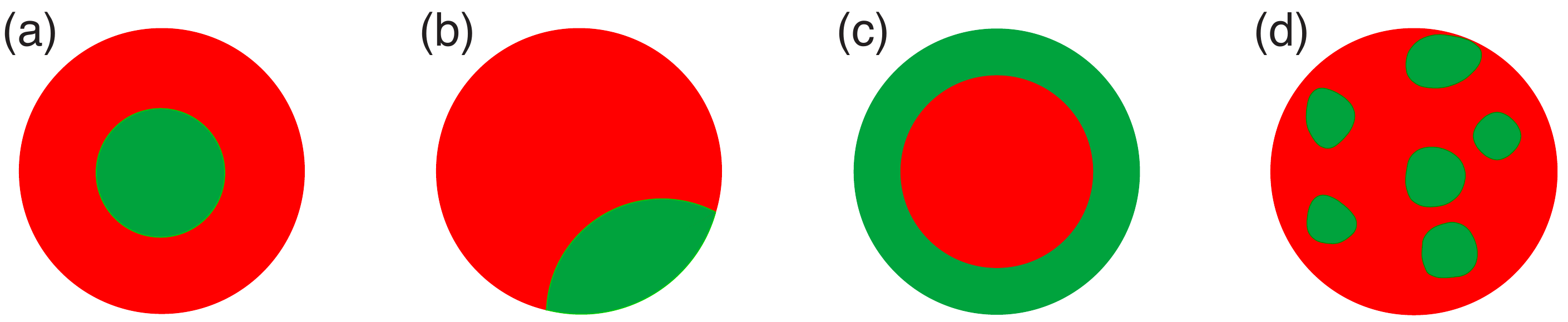}}
\caption{Domains patterns in the adhesion zone of T cells: (a) Final pattern of helper T cells with a central TCR domain (green) surrounded by an integrin domain (red) \cite{Monks98,Grakoui99}. The pattern results from cytoskeletal transport of TCRs towards the adhesion zone center \cite{Mossman05,Weikl04}. (b) Simulated final pattern in the absence of TCR transport \cite{Weikl04}. The length of the boundary line between the TCR and the integrin domain is minimal in this pattern. (c) and (d) The two types of intermediate patterns observed in the first minutes of adhesion \cite{Davis04}. In simulations, both patterns result from the nucleation of TCR clusters in the first seconds of adhesion and the subsequent influx of unbound TCR and MHC-peptide ligands into the adhesion zone \cite{Weikl04}. The closed TCR ring in pattern (c) forms from fast-growing TCR clusters in the periphery of the adhesion zone at sufficiently large TCR-MHC-peptide concentrations. The pattern (d) forms at smaller TCR-MHC-peptide concentrations.}
\label{figure_patterns}
\end{center}
\end{figure}

The domains of long and short receptor-ligand complexes formed during T cell adhesion evolve in characteristic patterns. The final domain pattern in the T-cell contact zone is formed within 15 to 30 minutes and consists of a central TCR domain surrounded by a ring-shaped integrin domain \cite{Monks98,Grakoui99} (see Fig.~\ref{figure_patterns}(a)). In contrast, the  intermediate patterns formed within the first minutes of T-cell adhesion are either the inverse of the final pattern, with a central integrin domain surrounded by a ring-shaped TCR domain (see fig.~\ref{figure_patterns}(c)), or exhibit several nearly circular TCR domains in the adhesion zone (see fig.~\ref{figure_patterns}(d)) \cite{Grakoui99,Davis04}. To understand this pattern evolution, several groups have modelled and simulated the time-dependent pattern formation during adhesion \cite{Qi01,Weikl02a,Burroughs02,Lee03,Raychaudhuri03,Weikl04,Coombs04,Figge06,Tsourkas07,Tsourkas08,Fenz17,Knezevic18}. Monte Carlo simulations with a discretized elastic-membrane model indicate that the central TCR cluster is only formed if TCR molecules are actively transported by the cytoskeleton towards the center of the adhesion zone \cite{Weikl04}. The active transport has been simulated by a biased diffusion of TCRs towards the adhesion zone center, which implies a weak coupling of TCRs to the cytoskeleton. In the absence of active TCR transport, the Monte Carlo simulations lead to the final, equilibrium pattern shown in Fig.~\ref{figure_patterns}(b), which minimizes the energy of the boundary line between the TCR and the integrin domain \cite{Weikl04}. In agreement with these simulations, T-cell adhesion experiments on patterned substrates indicate that cytoskeletal forces drive the TCRs towards the center of the adhesion zone \cite{Mossman05,DeMond08}, by a weak frictional coupling of the TCRs to the cytoskeletal flow initiated by T cell activation. The intermediate patterns formed in the Monte Carlo simulations closely resemble the intermediate T cell patterns shown in Figs.~\ref{figure_patterns}(c) and (d). In the simulations, these patterns emerge from small TCR clusters that are formed within the first seconds of adhesion \cite{Weikl04}. The diffusion of free TCR and MHC-peptide molecules into the adhesion zone leads to faster growth of TCR clusters close to the periphery of the adhesion zone. For sufficiently large TCR-MHC-peptide concentrations, the peripheral TCR clusters grow into the ring-shaped domain of Fig.~\ref{figure_patterns}(c). At smaller TCR-MHC-peptide concentration, the initial clusters evolve into the multifocal pattern of Fig.~\ref{figure_patterns}(d). 

Experiments on reconstituted membrane systems and simulations show that curvature-mediated segregation also occurs between small receptor-ligand complexes and larger membrane-anchored proteins that sterically prevent the preferred local separation of the receptors and ligands for binding \cite{Schmid16}. Calculations and simulations with discretized elastic-membrane models indicate segregation of sterically repulsive proteins of length $l_r$ and receptor-ligand complexes with preferred local separation $l_0< l_r$ if the concentration of the repulsive proteins exceeds the critical concentration $c_r \,k_B T/\kappa_\text{ef}(l_r -l_0)^2$ with a numerical prefactor $c_r$ \cite{Weikl02a}. The interplay of receptor-ligand binding and steric repulsion from glycocalyx proteins has also been investigated in models for the clustering of integrin receptors during cell adhesion \cite{Atilgan09,Paszek09,Paszek14,Xu16}. In addition, segregation of long and short bonds has been observed in experiments with reconstituted membranes that adhere {\em via} anchored DNA \cite{Chung13}. Segregation can also result from the interplay of specific receptor-ligand binding and generic adhesion if the minimum of the generic adhesion potential is located at membrane separations that are significantly larger than the lengths of the receptor-ligand complexes \cite{Weikl02b,Komura00,Fenz17}.

\subsection{Fluctuation-mediated attraction of adhesion complexes}
\label{section_fluctuation_mediated}

Complexes of receptor and ligand proteins that are anchored in apposing membranes `clamp' the membranes together and constrain the shape fluctuations of the membranes. These constraints lead to attractive fluctuation-induced interactions between identical receptor-ligand bonds, or receptor-ligand bonds of equal lengths. The interaction of two neighboring bonds is attractive because the membrane shape fluctuations are less constrained if the bonds are close to each other. The membranes then are effectively clamped together at a single site, and not at two sites as for larger bond distances. A single receptor-ligand bond that pins two fluctuating membranes locally together leads to a cone-shaped average membrane profile around the pinning site, i.e.\ the average local separation of the  membranes increases linearly with the distance from the receptor-ligand bond \cite{Bruinsma94,Netz97} because of the steric, fluctuation-induced  repulsion of the membranes \cite{Helfrich78,Lipowsky86}. Pair interactions of two isolated receptor-ligand bonds can be understood as interactions of two cone-shaped profiles around the bonds \cite{Farago10}, but this interaction appears artificial because membrane adhesion is typically mediated by many bonds, and the apposing membranes are on average planar and parallel to each other in adhesion zones with many bonds of the same type or length. As an alternative, pair interactions of receptor-ligand bonds can be calculated for membranes held together by an additional generic harmonic potential, but these pair interactions then depend on the potential strength and the location of the potential minimum \cite{Speck10}.

A more suitable approach is to quantify the overall strength of the fluctuation-mediated interactions between receptor-ligand bonds from the phase behavior and aggregation tendency of many bonds \cite{Weikl00,Weikl01}. The overall strength of the fluctuation-mediated attraction between identical receptor-ligand complexes depends on the length of the complexes, on the strength of the membrane confinement exerted by a complex, and on the concentration of the complexes. Because of the fluctuation-induced attraction, the receptor-ligand bonds in Fig.\ \ref{figure_multiscale_model}(b) tend to be in the vicinity of other receptor-ligand complexes. However, the fluctuation-mediated interactions are not sufficiently strong to induce aggregation and domain formation on their own,   irrespective of the confinement strength and concentration of the bonds \cite{Weikl01,Weikl06}. Aggregation only due to fluctuation-mediated interactions does not occur in the elastic-membrane model of Fig.\ \ref{figure_elastic_membranes}(a) because the unbinding transition of membranes in the effective adhesion potential (\ref{Vef}) of this model is continuous \cite{Lipowsky86}, which precludes a discontinuity in concentrations during adhesion that is necessary for aggregation and domain formation \cite{Weikl01,Weikl06}. Bond aggregation either requires additional direct attractive interactions \cite{Weikl01,Weil10}, an additional local stiffening of the membranes by the membrane anchors of the receptors or ligands \cite{Weikl01}, or a preclustering into multimeric receptors and ligands \cite{Weikl00}.

\section{Summary and outlook}

Biomembrane adhesion involves an interplay between protein binding and membrane shape fluctuations, because the binding of the proteins that mediate the adhesion depends on the separation of the two adhering membranes. The elastic-membrane and coarse-grained molecular models of biomembrane adhesion reviewed here indicate that the binding equilibrium resulting from this interplay can be understood based on Eq.\ (\ref{K2Dav}) for the apparent binding constant $K_\text{2D}$ of the proteins (see section \ref{section_K2D}). The apparent binding constant $K_\text{2D}$ depends on properties of the protein molecules as well as on properties of the membranes and is therefore not a proper binding constant. The proper quantification of the binding affinity of proteins in membrane adhesion is the function $k_\text{2D}(l)$ in Eq.\ (\ref{K2Dav}), i.e.\ the binding constant of the proteins as a function of the local membrane separation $l$. The function $k_\text{2D}(l)$ depends only on molecular properties of the proteins, i.e.\ on their membrane anchoring, lengths, flexibility, and binding interaction, and is thus an equivalent of the binding constant $K_\text{3D}$ of soluble proteins. The dependence of $K_\text{2D}$ on overall properties of the membranes is captured by the distribution $P(l)$ of local membrane separations in Eq.\ (\ref{K2Dav}). The distribution $P(l)$ depends on the concentration of the protein complexes because the protein complexes constrain the membrane shape fluctuations, which leads to a binding cooperativity (see section \ref{section_cooperative_binding}). Membrane shape fluctuations also lead to attractive interactions between receptor-ligand bonds of similar lengths (see section \ref{section_fluctuation_mediated}), besides cooperative binding. However, these fluctuation-mediated interactions are not sufficiently strong to induce domains of receptor-ligand bonds on their own. In contrast, the curvature-mediated repulsion between  receptor-ligand bonds of different lengths is a strong driving force for segregation and domain formation (see section \ref{section_length_based}). This curvature-mediated repulsion results from the membrane curvature required to compensate the length mismatch between short protein bonds and long bonds or proteins. 

A future challenge of simulations and experiments is to determine the binding constant function $k_\text{2D}(l)$ of membrane-anchored proteins. For the CD47-SIRP$\alpha$ complex of Fig.\ \ref{figure_multiscale_model}(a), the shape of the function $k_\text{2D}(l)$ has been obtained from simulations with a coarse-grained molecular model of the bound complex (see Fig.\ \ref{figure_CD47_data}(b)). In principle, values of $k_\text{2D}(l)$ at selected local separations $l$ can be obtained  from modelling and simulations by determining the probability that the membrane-anchored proteins are bound at these local separations. A promising approach is to focus on the ratio $k_\text{2D}(l)/K_\text{3D}$ because this ratio is determined by the molecular architecture and anchoring flexibility of the membrane-anchored proteins, which can be well described with coarse-grained modelling (see also Eq.\ (\ref{k2DoverK3D}) for rod-like receptors and ligands). The free-energy contribution of the binding site drops out from this ratio because the binding site can be assumed to be identical for the soluble and membrane-anchored proteins. Absolute values for $k_\text{2D}(l)$ can then be obtained if the binding constant $K_\text{3D}$ of soluble variants of the proteins without membrane anchors is known from experiments. A further experimental challenge is to measure the relative roughness of the adhering membranes. Experimental measurements of the relative membrane roughness require a spatial resolution in the nanometer range both in the directions parallel and perpendicular to the membranes, which is beyond the scope of current methods \cite{Pierres08,Lin14,Monzel15}.

\section*{Acknowledgements}

We would like to thank Mesfin Asfaw, Heinrich Krobath, Emanuel Schneck, Guang-Kui Xu, and in particular Reinhard Lipowsky
for numerous discussions and joint work on topics of this review. 

BR has been supported by the National Science Centre, Poland, grant number 2016/21/B/NZ1/00006. JH has been supported by the National Science Foundation of China, grant numbers 21504038 and 21973040. Part of the simulations have been performed on the computing facilities in the High Performance Computing Center (HPCC) of Nanjing University.

\small

\end{document}